# The Structure of Hydrated Electron.

# Part 1. Magnetic Resonance of Internally Trapping Model Water Anions: A Density Functional Theory Study.

Ilya A. Shkrob

*Chemistry Division , Argonne National Laboratory, 9700 S. Cass Ave,*

*Argonne, IL 60439*

## Abstract

Density functional theory (DFT) is used to rationalize magnetic parameters of hydrated electron trapped in alkaline glasses as observed using Electron Paramagnetic Resonance (EPR) and Electron Spin Echo Envelope Modulation (ESEEM) spectroscopies. To this end, model water cluster anions ($n$=4-8 and $n$=20,24) that localize the electron internally are examined. It is shown that EPR parameters of such water anions (such as hyperfine coupling tensors of H/D nuclei in the water molecules) are defined mainly by the cavity size and the coordination number of the electron; the water molecules in the second solvation shell play a relatively minor role. An idealized model of hydrated electron (that is usually attributed to L. Kevan) in which six hydroxyl groups arranged in an octahedral pattern point towards the common center is shown to provide the closest match to the experimental parameters, such as isotropic and anisotropic hyperfine coupling constants for the protons (estimated from ESEEM), the second moment of the EPR spectra, and the radius of gyration. The salient feature of these DFT models is the significant transfer (10-20%) of spin density into the frontal $O\ 2p$ orbitals of water molecules. Spin bond polarization involving these oxygen orbitals accounts for small, negative hyperfine coupling constants for protons in hydroxyl groups that form the electron-trapping cavity. In Part 2, these results are generalized for more realistic geometries of core anions obtained using a dynamic one-electron mixed qunatum/classical molecular dynamics model.

---

[a] Author to whom correspondence should be addressed; electronic mail: shkrob@anl.gov.

## 1. Introduction.

This paper continues a series of publications [1-4] on the "bottom up" approach to the structure of solvated/trapped electrons in molecular liquids and glasses. In this two-part series we consider the most important species of this kind: the hydrated electron, $e_{hyd}^-$. [5] Closing a 30 year gap in the literature, we revisit the magnetic properties of the electron trapped in alkaline ice and compare *ab initio* and density functional theory (DFT) calculations of such properties for gas phase $(H_2O)_n^-$ ($n$=4-24) clusters and hyperfine coupling (hfcc) tensors for magnetic nuclei that were determined experimentally in the 1970s and the 1980s. While this comparison upholds several commonly assumed features for the cavity model of $e_{hyd}^-$, it also suggests that one-electron theories of the hydrated electron might be incomplete. The salient feature that is missing from these theories is the significant transfer (10-20%) of the spin density into the frontier O $2p$ orbitals of water molecules forming the solvation cavity. There have been recent suggestions [6] that this transfer might account for the observed 200 cm$^{-1}$ downshift of the O-H stretch mode and ca. 30 cm$^{-1}$ downshift H-O-H bend modes in the resonance Raman spectra of $e_{hyd}^-$ in liquid water. The examination given in this paper suggests that the magnetic resonance properties of $e_{hyd}^-$ cannot be understood in any other way. In Part 2 of this series, [7] water configurations generated by a mixed quantum /classical molecular dynamics (MQC MD) model are used as the input for the embedded cluster DFT calculation to generate the statistically averaged picture of $e_{hyd}^-$ in liquid water. This hybrid approach reproduces several heretofore unexplained properties of $e_{hyd}^-$, including its spin parameters and IR-Raman features. While the examination in the present paper does not address the variability of the core anion structure and the effect of the solvent beyond the second solvation shell, the general conclusions reached here are upheld in Part 2 of this series. [7]

In water and aliphatic alcohols the excess electron is stabilized via strong electrostatic interaction with non hydrogen bonded (NHB) hydroxyl groups pointing towards the solvation cavity filled by the electron wavefunction (Figure 1). Pauli exclusion and core repulsion stabilize this cavity against the collapse. Traditionally, solvated electrons were treated using one-electron models in which the excess electron is considered separately from the valence electrons in the solvent (which is described classically). In these models, the electron interacts with the solvent molecules by means of an *ad hoc* empirical, classical potential. This approach has been introduced in the mid 1950s, [5] and it has been immensely successful. Since the 1980s, hydrated electron became the test bed for state-of-the-art path integral [8] and MQC MD [9-12] calculations in which the solvent motion is treated dynamically, at the classical level, whereas a single quantum mechanical particle, the excess electron, is treated quantum mechanically, in the adiabatic [9,12] or nonadiabatic [9,10] approximations. The MQC MD approach allows one to treat the electron relaxation, pump-probe dynamics, excitation spectrum, etc. straightforwardly, with relatively few further assumptions. This approach proved to be both insightful and productive, and it greatly refined the theoretical picture of electron solvation. However, despite these many successes, the one-electron models, regardless of their technical sophistication, suffer from two closely related problems:

One of these problems is that of justification. It does not follow from any higher-level theory or a general principle that the one-electron models in which the solvent is described classically and a single electron is described quantum mechanically is the adequate picture of $e_{hyd}^-$. The reasoning goes the other way round. The one-electron model is *postulated* and then the consequences of this assumption are tested against the experiment. Good agreement with the experiment is then taken as the justification for the assumptions and simplifications introduced in the model. The pitfall of this approach is that more than one theory is capable of accounting for a given group of experimental observations, especially when empirical $e^- - H_2O$ pseudopotentials are allowed. The majority of theoretical studies of $e_{hyd}^-$ focused on a single property: the absorption spectrum in the visible. Despite great differences in the approach and the degree of detail, all of 250+ theoretical papers on $e_{hyd}^-$ claim good understanding of this spectrum and its salient features. A possible conclusion is that this absorption spectrum might not be too revealing about the details of electron solvation, once the model satisfies a few rather general criteria. (In this regard, the ability of MCQ MD calculations [9-12] to correctly reproduce the basic features of pump-probe experiments on $e_{hyd}^-$ is more important than the explanation of the static spectrum). What is more troubling, similar one-electron models have been used [13] (with the same degree of fidelity) to account for the absorption spectra of solvated electron in liquid ammonia and aliphatic amines, for which there is a strong case, supported both by theoretical considerations [1,14,15] and nuclear magnetic (NMR) [16,17] and electron paramagnetic resonance (EPR) spectroscopies (reviewed in ref. 1), that the one-electron picture is incorrect, as most of the spin density is contained in the forward N $2p$ orbitals of the solvent molecules in the first and the second solvation shells. [1,16] In other words, the one-electron models, even when these are demonstrably incorrect in their basic assumptions, still account quite well for the optical properties of the excess electron. Such observations bring to the fore the question of how justifiable such one-electron models are in general. [1-4]

One way to justify and support these models would be calculating the less frequently addressed properties of $e_{hyd}^-$. There lies the second pitfall of the one-electron models since by their very nature these are not conducive to such tests. In particular, the two experimental methods that give the most direct and detailed structural information on the ground state of the solvated/trapped electron – magnetic resonance data for trapped electrons in alkaline ices (section 2) and resonance Raman spectroscopy of hydrated electrons in liquid water [6] – are the least tractable from the standpoint of these theories, as the solvent molecules can no longer be considered classically. Other properties of $e_{hyd}^-$ that do not lend themselves easily to such calculations are its vacuum UV band at 190 nm [18] and the proton transfer reactions. [5] The 190 nm band supposedly involves O $2p$ orbitals in the water molecules forming the cavity, [18] whereas the latter requires full quantum mechanical treatment of water molecules. A limited set of experimental results is revisited repeatedly, whereas other equally important properties of $e_{hyd}^-$ remain seldom addressed.

Over the last decade, this situation has changed, largely due to the advances in anion cluster chemistry. The need for understanding the properties of gas phase $(H_2O)_n^-$ anions [19] fomented interest in modeling such species using *ab initio* and DFT methods that go beyond the one-electron approximation.[20,21,22] Such calculations were originally carried out for relatively small clusters ($n$=6-10) that trap the electron at their surface, yielding dipole-bound anions. The internally solvated electrons can also be modeled using such small clusters,[22] but their structure does not correspond to any known species observed experimentally in the cluster studies. Recently, the increased computational power allowed to examine several larger clusters ($n$=20 and 24) that demonstrate internal solvation by four NHB[20,21] hydroxyl groups (while most of the remaining OH groups are involved in the H-bond formation). Another promising development was the recent Car-Parrinello MD modeling[23] of hydrated electron in bulk water, although the only experimentally property calculated was again the optical spectrum in the visible.

Most of these studies focused on the energetics of the water anion clusters. Yet the approach also allows to estimate the magnetic parameters for the $^1$H and $^{17}$O nuclei in the water molecules and compare these estimates with EPR parameters for trapped electron in alkaline ice. Such is the program implemented in this paper. Before summing up these calculations, the results of EPR, Electron Spin Echo Envelope Modulation (ESEEM), $^1$H Electron Nuclear DOuble Resonance (ENDOR) and ELectron-electron DOuble Resonance (ELDOR) spectroscopy of trapped electrons in alkaline ice are reviewed. The bulk of these results was obtained in the mid-1970s. The initial studies culminated in the well-known octahedral model of $e_{hyd}^-$ (Figure 1(a)) that is commonly associated with the name of L. Kevan,[24] although it was first suggested by Natori and Watanabe[25a] and Natori.[25b] In this model, the cavity is formed by six NHB hydroxyl groups pointing towards the common center. In fact, neither Kevan's EPR, ENDOR, and ESEEM studies[24,26-30] nor the follow-up ESEEM studies pursued by Bowman's and Tsvetkov's groups[31,32,33] lend support to this model (section 2). Surprisingly, the first solid evidence that this model does capture, albeit approximately, some of the properties of hydrated electron that are relevant for magnetic resonance is provided by this study. Since few theorists currently involved in modeling $e_{hyd}^-$ and $\left(H_2O\right)_n^-$ clusters are familiar with magnetic resonance studies carried out 20-30 years ago, the methods used, and the controversies involved, the subject is briefly re-examined in section 2. The basics of EPR and ESEEM spectroscopies pertaining to the discussion in section 2 are discussed in Appendix A in the Supplement.

**2. Magnetic Resonance Studies.**

Though water can be prepared in a glassy state via hyperquenching of aerosol droplets, radiolysis of such glassy water does not yield trapped electrons, [34a] presumably due to the high concentration of proton defects that react with the electrons. Without salt, water crystallizes into hexagonal $I_h$ ice. Upon radiolysis, low-temperature $I_h$ ice yields two kinds of trapped-electron centers. [34b] Electron-1 is short-lived (<1 μs) and absorbs in the visible; electron-2 (that is observed only below 40 K and only in $D_2O$ ice) is long lived (slowly decaying by tunneling to the hole centers) and absorbs in the near IR. The two species do not interconvert, even after photoexcitation. The EPR spectrum of electron-1 is unknown due to its short lifetime; the EPR spectrum of electron-2 at 4.2 K (with peak-to-peak width $\Delta B_{pp}$ of the EPR line of 1.3 G; 1 G = $10^{-4}$ T) has been reported by Johnson and Moulton [34c] and Hase and Kawabata. [34d] In $H_2O$ ice this line width would correspond to 8.3 G (the magnetic moment of the deuteron is 15.4% of that for the proton). For a Gaussian line the second moment of the spectral line $M_2 = \Delta B_{pp}^2 / 4$, which gives $M_2\left(^1H\right) \approx 17.5$ G$^2$ vs. 21-23 G$^2$ for trapped electron in alkaline ices. Electron-1 is believed to be trapped by a Bjerrum proton disorder defect in ice, [34b,35] whereas electron-2 is believed to occupy a relatively undistorted hexagonal cavity. [36] Neither of these attributions have been proved.

Since water is not a natural glass former and the IR absorbing trapped electron species in $I_h$ ice have no apparent connection to $e_{hyd}^-$ in liquid water, the EPR studies of hydrated electrons were primarily conducted on alkaline ice glasses containing 5-15 M of alkali hydroxides (typically, NaOH). The first such EPR spectrum was obtained by Bennet et al. by deposition of Na atoms on polycrystalline ice; [37] that was followed by studies on γ-irradiated alkaline glasses containing a relatively low fraction of NaOH. [38,39] Such glasses include numerous microscopic ice crystallites, whose fraction, size, and morphology vary from sample to sample. These crystallite inclusions seemed to improve the spectral resolution. Poorly resolved, poorly reproducible septets of resonance lines were reported and estimates of ca. 5-6 G for the isotropic hyperfine couplings constants (hfcc's) $a^H$ for the protons were obtained (5.7 G [27] from ELDOR, and 6 G, [38] 5.6 G, [37] and 4.7 G [39] from EPR). The octahedral model was suggested by Natori et al. [25] to account for this 7-line pattern. No such spectra have been observed since 1972; the subsequent studies of fully glassified samples yielded featureless EPR spectra. Although the original justification for the octahedral model [24,25] was this irreproducible 7-line EPR spectrum, the model stubbornly persists to this day.

The goal of the EPR [40] and ESEEM [33] studies is to provide hyperfine coupling tensors for $^1$H (or $^2$H) nuclei in the hydroxyl groups lining the solvation cavity. These tensors can be used to map the electron wavefunction and determine (within certain approximations) the geometry of the solvation cavity. The hyperfine coupling tensor **A** with principal values of $(A_{xx}, A_{yy}, A_{zz})$ can be represented as $\left(a + B_{xx}, a + B_{yy}, a + B_{zz}\right)$, where $a$ is the isotropic hyperfine coupling constant (originating through the Fermi contact interaction) and **B** is the traceless tensor of anisotropic hyperfine interaction that originates through electron-nuclear magnetic dipole coupling. Typically, such tensors are nearly axial, so that $B_{xx} \approx B_{yy} \approx T_\perp$ and $B_{zz} \approx -2T_\perp$. For a point like electron interacting with a nucleus at a distance $r$, $T_\perp = \gamma_e \gamma_n / h r^3$, where $\gamma_{e,n}$ are the corresponding gyromagnetic ratios and $h$ is the Plank constant. For a proton, $B_{zz}^H \approx 57.6 / r^3$, where the distance is given in Å. All of the EPR and ESEEM data for trapped electrons were interpreted using this point dipole approximation, although it is not obvious that the latter holds for a cavity occupied by spatially extended electron wavefunction. Nevertheless, all $r_{XH}$ distances (where $H$ denotes the nearest proton and $X$ is the centroid of the electron density in the cavity) were estimated from the experimentally determined $T_\perp$ using this point dipole approximation. Only in the retrospect was it realized by Golden and Tuttle [41] that this approximation might result in grossly incorrect values for $r_{XH}$ when the latter is in the range of 2.1-2.4 Å (i.e., in the range suggested by the ESEEM data). The situation is further complicated when there is nonzero spin density on O atoms, as the protons would also interact with the unpaired electron in the O *2p* orbitals. This interaction decreases both the dipolar contribution and the isotropic constant (it can even reverse the sign of $a^H$, via spin bond polarization (see Appendix A). [1,14,15,40]

When there are many magnetic nuclei coupled to the electron, the EPR line becomes Gaussian and the second moment $M_2$ of this line is given by eq. (A7). In alkaline glasses with more than 5 M hydroxide, the peak-to-peak width of the EPR line linearly increases with the concentration of the alkali (KOH, NaOH, or CsOH), [32] and extrapolates to $9.5 \pm 0.5$ G at zero concentration (corresponding to $M_2 \approx 22.6$ $G^2$). Bales et al. [28,29] gave several other estimates, e.g. 21 $G^2$, [29] as there are slight sample-to-sample variations. The line broadening that increases with the increasing alkali concentration is due to the interaction of the electron with magnetic alkali nuclei ($^{40}$K, $^{23}$Na, and $^{133}$Cs) in the matrix. This poses a question whether the electron trapped in alkali glasses includes the alkali cation in its solvation shell. The original ESEEM study by Narayana et al. [24] suggested that the interaction with the (matrix) alkali cations are long range and dipolar in origin, and that the cations can therefore be excluded from consideration. However, subsequent ESEEM studies showed that suggestion to be incorrect. [31,33] Kevan's ESEEM experiments [24] were carried out on NaOD/D$_2$O samples, for which the spin echo modulation pattern induced by $^{23}$Na is masked by strong modulation by deuterons. In protiated glasses, this modulation is very fast, and the echo modulation by alkali nuclei is clearly observed. [31] Subsequent analyses [33] indicated that one of the $^{23}$Na nuclei (in 10 M NaOH/H$_2$O glass) is strongly coupled to the electron, with $a \approx +0.6$ MHz and $T_\perp \approx -1.2$ MHz (1 G = 2.8 MHz); the latter corresponds to $r_{XNa} \approx 2.6$ Å. Thus, the trapped electron in alkaline glasses does include the alkali cation in its first solvation shell. The same is suggested by the fact that the absorption band of the electron in alkali glasses and concentrated alkali solutions [42] is strongly blue shifted with respect to the band observed in dilute solutions. This suggests strong electrostatic interaction in a close pair. Just such an interaction was indeed observed theoretically in MQC MD models of hydrated electron by Boutin and co-workers. [43] These observations suggest that the electron trapped in the alkali glasses is not the same species as $e_{hyd}^-$ in liquid water.

Only hfcc tensors for the protons in the first solvation shell are of import for EPR and ESEEM studies. The contribution from remote (matrix) protons to the second moment $M_2^H$ is given by eq. (A10). For a cutoff radius of $r_{cut}$ the contribution of matrix protons to $M_2$ is given by $46.3\ G^2/r_{cut}^3$, which gives $1.4\ G^2$ (or $1.1\ G^2$) for the cutoff radius of 3.2 Å (or 3.5 Å [31] or 3.6 Å [24]) that corresponds to the next nearest protons. This immediately suggests that the main contribution to $M_2^H$ is provided by the closest NHB protons; thus the second moment can be used to constrain the cavity geometry. Additional constraints are suggested by other methods:

The narrow resonance line of the electron exhibits two partially resolved side lines, which are due to forbidden transitions involving the simultaneous flip of the electron and nuclear spins ("flip-flop satellites"). [29] From the satellite intensity (eq. (A11)), the second moments of the satellite, and the main lines in 10 M NaOH/$H_2O$ glass it is possible to obtain a crude estimate of the absolute $a$ and $T_\perp$ for the first shell protons if the mean number $\langle n \rangle$ of the (magnetically equivalent) protons is known. For $\langle n \rangle$ =4, 6, and 8 the following estimates were obtained for $|a|$ and $|2T_\perp|$: 2±3 G and 8.3±1 G (1.98 Å), 1±3 G and 6.8±0.8 G (2.12 Å), 1±2 and 5.9±0.8 G (2.22 Å, in the point dipole approximation), respectively. The estimates from $^1$H ENDOR line width also give $r_{XH} \approx 2.3 \pm 0.1$ Å. [26] The experiment of Narayana et al. [24] was interpreted assuming "Kevan's model:" six magnetically equivalent protons with $|a^D| \approx 0.9$ MHz (i.e., $a^H \approx 2.1$ G for $^1$H) at $r_{XH} \approx 2.1$ Å (the authors claimed that $a^D > 0$). The fit quality, however, was poor, and the fidelity of this model questionable. One of the reason for that was that at that time (1975) the theory of ESEEM was not sufficiently advanced to include weak nuclear quadrupolar interaction for spin-1 deuterons, which is important for fitting ESEEM kinetics. [33] The methods currently used for estimation the number of coupled nuclei involved in the formation of the ESEEM pattern were not yet developed, and the suppression of the matrix signals was not used in the analyses. [33] In 1988, Astashkin et al. [32] revisited the NaOD/$D_2O$ system and reanalyzed the echo kinetics using greatly improved methods. Their results contradicted the original findings of Narayana et al. [24] Astashkin et al. [32] obtained $a \approx \pm 0.4$ MHz (which is equivalent to ±0.92 G for $^1$H) and $T_\perp \approx \pm 1.5$ MHz ($r_{XH} \approx 2.01$ Å; $2T_\perp \approx \pm 7$ G for [1]H). Assuming that the perpendicular component $T_\perp$ of the hfcc tensor for the protons is negative, as expected in the point dipole approximation, the isotropic constant on the protons should also be *negative*. (Observe that such a small negative constant is still in accord with satellite line – second moment analyses of Bales et al.) [28,29] The possibility of a negative constant $a^H$ was originally dismissed because all one-electron models suggested so far (including semicontinuum electron models developed by Kevan and co-workers) [44] give only positively valued isotropic hfcc's for [1]H nuclei (as the negatively valued constants originate through spin bond polarization involving unpaired electrons in the O *2p* orbitals). Negatively valued isotropic hfcc constants (that were also observed, using NMR and dynamic nuclear polarization, for [1]H protons in ammoniated electron) [1,16,17] imply *the breakdown of the one-electron approach*. The same is also suggested by the anomalously small absolute value of the hfcc constant. In the absence of spin bond polarization, large positively valued estimates for $a^H$ (ca. 3-5 G) were suggested by these semicontinuum models. [25,44] Since the initial EPR experiments [37,38,39] seemed to yield such large hfcc constants (see above), these experiments were considered to be supportive of such one-electron models. The subsequent ESEEM experiments, [32,33] however, yielded hfcc estimates that are clearly incompatible with these one-electron models. By contrast, the early tight-binding *ab initio* calculations for water tetramer anions [14,15,45] yielded *negative* proton constants, suggesting significant spin bond polarization in such water anions. This discrepancy was acknowledged by Kevan and co-workers [30] but considered unimportant in the view of their ESEEM results. [24] As seen from the above, subsequent ESEEM studies hinted at just such a polarization.

One of the detriments of ESEEM spectroscopy [33] for species that exhibit structural disorder (such as trapped electrons in glasses) is that it is difficult to conclude how many nuclei are contributing to the observed modulation pattern, as the latter originates through averaging over many such nuclei, including those of the matrix (i.e., distant ones). The original conclusion that the coordination number is six [24] was based on poor quality kinetic simulation using (as was belatedly realized) [32,33] incorrect hfcc parameters. The more recent simulations of Astashkin et al. [32] suggested a different picture: the first solvation shell is comprised of only *two* protons at $r_{XH} \approx 2$ Å from HO groups pointing towards the cavity; in addition, there are 7-8 other water molecules with mean $r_{XH} \approx 3.5$ Å that show no preferential orientation of these OH groups. MQC MD [9-12] and path integral [8] (one electron) models of $e_{hyd}^-$ generally yield the first peak in the X-H distribution at 2 Å and a coordination number that is close to six. While the distances reported by Astashkin et al. [32] broadly agree with these MQC MD estimates, their estimate for the coordination number is suspect since the predicted second moment $M_2$ in protiated alkaline glass [32] is 40-50% lower than observed. [28,29] Simulating the EPR spectrum of $e_{hyd}^-$ using this 2-deutron model yields a 7-line pattern resembling the one observed by Ohno et al. [39] in 2-5 M alkaline glass, but it does not resemble (neither in its shape nor in the envelope width) the EPR spectrum observed in 10 M glasses for which the ESEEM spectra were obtained. [24,32,33]

Similar methods were used to obtain estimates for hfcc parameters for electrons trapped in other polar glasses. According to Kevan et al, [46] for electrons in ethanol glass, the mean coordination number $\langle n \rangle$, the X-H distance $r_{XH}$, and the isotropic hfcc $|a|$ for deuterons in the methyl, methylene, and hydroxyl groups are 12, 3.8 Å, and 0 MHz ($CD_3$), 8, 3.3 Å, and 0.1 MHz ($CD_2$) and 4, 2.2 Å, and 0.7 MHz (OD). In isotope substituted ethylene glycol glass, [47] the electron is coordinated by two nonequivalent OD groups (2.7 Å and $|a|$=0.3 MHz) and 16 methylene deuterons (four at 3 Å and twelve more at 4.2 Å); there are four ethylene glycol molecules in the first solvation shell, but only two of these contribute their OH bonds. Thus, it appears that in the alcohol glasses, there are four rather than six NHB protons per cavity. Subsequently it has been discovered that even crystalline carbohydrates can trap electrons at low temperature. [48] Due to a high degree of structural order, it is possible to obtain detailed information on the geometry of the electron trapping center in the irradiated sugar crystals using [1]H ENDOR. In all of these sugars with exception of arabinose, the electron is trapped either by *two* NHB protons or two *sets* of magnetically equivalent NHB protons that are located at distances ranging from 1.6 Å to 1.82 Å (see Table 1 in ref. 48); the absorption band of the electron in such crystals is located to the blue of trapped electron in alkaline glasses, suggesting a deep trap. To this day, all attempts to reconcile these structural data with the crystallographic data and the observed electron energetics using semicontinuum models have failed. [48]

Another EPR observation that played a significant role in the universal acceptance of one-electron models in the mid-1970s was the experiment of Schlick at al. [30] in which the second moment for a 10 M NaOH/$H_2O$ sample enriched by 37 at% $^{17}O$ was obtained (unlike $^{16}O$, oxygen-17 is a magnetic, spin-5/2 nucleus). The contribution $M_2\left(^{17}O\right)$ from oxygen-17 was estimated at 134 $G^2$ (for $\Delta B_{pp} \approx 18\pm1$ G) which is equivalent to 362 $G^2$ for isotopically pure water sample. Schlick et al. [30] correctly assumed that (contrary to the situation with the proton contribution to $M_2$), the contribution from oxygen-17 is dominated by *isotropic* component of the hfcc tensor. Assuming $sp^3$ hybridization of the oxygen orbital, they estimated that for $\langle n \rangle$ =4, 6, and 8, $a\left(^{17}O\right)$ is ca. –5.6, -4.5, and –4.0 G (the negative sign is due to the negative nuclear moment of $^{17}O$); the corresponding $\left|2T_\perp\right|$ was estimated as 2 G. The corresponding isotropic and anisotropic hfcc constants for the atomic O *2p* orbital are $\left|a^O(at.)\right|$ =1660 G and $\left|B_{zz}^0(at.)\right|$ =102-104 G, respectively, [40] which suggested that the total transfer of the spin density into the O *2s* and *2p* orbitals is ca. 2% and 10-16%, respectively. As shown in section 4, the latter estimate appears to be correct; furthermore, our DFT calculations indicate that such a population would result in the observed negative isotropic hfcc's for NHB protons. The problem, however, is that subsequent EPR studies by Schlick and Kevan [49] put their initial result in doubt, as it was realized that the EPR signal from the electron overlapped with an inhomogeneously broadened $g_\parallel$ component (*m*=0) from the $^{17}O^-$ radical (that is formed in radiolyzed alkali glasses [49] and low temperature $I_h$ ice [50] by deprotonation of OH radical). In the oxygen-16 sample, the narrow lines from the O⁻ radical anion and the electron are spectrally separated, but in the oxygen-17 sample these two sets of resonance lines overlap. Since the resonance line of the electron is very saturable, it is difficult to correctly determine its line width in such a complex situation, especially as there is also a large contribution from (the majority) oxygen-16 sites. For that reason, the measurement was unreliable. The calculations given below suggest that under no realistic situation would the second moment from oxygen-17 nuclei be as low as 360 $G^2$; the estimates are at least an order of the magnitude greater. In Part 2, [7] we suggest that the narrow resonance line of the electron in oxygen-17 substituted sample was from isotopic configurations in which $^{17}O$ nuclei are in the second rather than the first solvation shell. The latter yield extremely broad EPR lines that are difficult to observe using EPR, especially in the presence of many other lines.

To conclude this section, magnetic resonance experiments carried out in the 1970s and 1980s did not give *direct* structural insight into the trapped electron in alkaline glasses. Rather, several important constraints on the electron structure were obtained and proton/deuteron hfcc tensors were estimated. The interpretation of what these results reveal about the structure of the electron depends on the theoretical model. For example, the estimates of cavity size depend on the validity of the point dipole approximation, the applicability of which is not obvious and, in fact, has been questioned. [41] Due to the lack of adequate models, no interpretation of the isotropic hfcc's consistent with the experimental estimates has been suggested. The EPR and ESEEM data indicate that such a model must be multi- rather than one- electron, given the small, negative isotropic constants for NHB protons (indicative of significant spin bond polarization). Thus, far from being able to provide direct structural information, the magnetic resonance results themselves need to be understood and interpreted in a self-consistent way using advanced theoretical models. Such a program is implemented below.

To reduce the length of the paper, the sections, tables, and figures with the designator "S" (e.g., Figure 1S) are placed in the Supplement.

### 3. Computational Details.

In this study, gas phase water cluster anions were analyzed mainly using density functional theory (DFT) models with B3LYP functional (Becke's exchange functional [51] and the correlation functional of Lee, Yang, and Parr) [52] from Gaussian 98. [53] B-LYP functionals are most frequently used to estimate isotropic hfcc in radicals and radical ions, for which it typically yields accurate and reliable results. As a complementary approach, self-consistent field Hartree-Fock (HF) and second-order MØller-Plesset (MP2) perturbation theory [54] calculations were used. The latter two methods gave very similar estimates for *magnetic* parameters, so in most cases only HF results are discussed below. While the anisotropic hfcc's calculated using these DFT and *ab initio* methods were comparable, the isotropic hfcc's differed substantially: the HF and MP2 generally yields smaller absolute isotropic hfcc ($a^O$) for $^{17}O$ nuclei and larger isotropic hfcc ($a^H$) for the innermost $^1H$ nuclei as compared to DFT methods (such as B-LYP and LSDA). This difference can be traced to the fact that the DFT models better account for the spin bond polarization effects (which accounts for their preferred use for the calculations of EPR parameters).

Unless specified otherwise, the basis set was a 6-31G split-valence double-$\zeta$ Gaussian basis set augmented with diffuse and polarized (d,p) functions (6-311++G**). Very similar results were obtained using two other basis sets, augmented Dunning's correlation consistent double basis set (aug-cc-pVDZ)[55] and Barone's triple-$\zeta$ basis set[56] with diffuse functions and an improved *s*-part that was introduced specifically for the hfcc calculations (EPR-III). Reduction of the basis set to 6-31+G** or smaller sets gave rather different results from those obtained using these basis sets (see, for example, Tables 1S and 4S). That was not the case in ammonia clusters examined in our previous study.[1] This is because in water anion clusters, the spin density inside the cavity is substantially greater than in the ammonia clusters and more diffuse sets are required to obtained reliable estimates. This is an important point, because the early *ab initio* studies [14,15] of water tetramer anions related to EPR used tight 3-21G and 4-31G basis sets. We also used basis sets (6-31+G** sets complemented by diffuse functions for hydrogen and oxygen atoms) that were developed by Bradforth and Jungwirth[57] and Herbert and Head-Gordon[21] for *ab initio* modeling of water anion clusters. The hfcc tensors obtained using these basis sets were very similar to those obtained using the standard 6-311++G** set.

In most of these calculations, a ghost atom (i.e., floating-center set of diffuse functions) at the center of the cluster was added with parameters used in refs. 21 and 57. For the 6-311++G** set and other large basis sets, the introduction of this ghost atom had little effect on the calculated hfcc tensors.

Three types of the model clusters were examined: (i) small, highly symmetrical $(H_2O)_n^-$ clusters ($n = 4$, 6, and 8) in which water molecules were arranged in such a way that the hydroxyl group of each molecule pointed towards the common center $X$ ($b$-type clusters), or with the water dipoles pointing to the same center ($d$-type clusters), (ii) four $n = 20$ and 24 clusters that internally trap electrons (the cluster geometries were obtained from Khan [20] and Herbert and Head-Gordon; [21] the geometries of these clusters are given in Appendix B in the supplement), and (iii) embedded clusters generated from 200 snapshots of the hydrated electron obtained in the MQC MD calculation (Part 2).

In addition to the hfcc tensors, second moments $M_2^H$ and $M_2^O$ were calculated for [1]H and [17]O nuclei, respectively, using eq. (A7) (the contributions from isotropic and anisotropic parts of the hfcc tensor were calculated separately). We also used these hfcc parameters and the directional cosines for hfcc tensors to directly simulate EPR spectrum for randomly oriented fixed-geometry clusters (assuming a spherical $g$-tensor). For small, highly symmetrical clusters the simulated EPR spectra exhibited some structure, but the spectra simulated for $n=20$ and $n=24$ anion clusters (both for [1]H[16]O and [1]H[17]O) are nearly Gaussian and show no spectrally resolved resonance lines, like the experimental spectra obtained in alkaline glasses. Typical examples of such spectra are shown in Figure A1(a) in the Supplement.

Since in a multi-electron model of the solvated/trapped electron the definition of what constitutes the cavity is ambiguous, it is difficult to quantify the partition of the spin density between the cavity and the solvent molecules exactly. Examination of density maps for the spin-bearing highest occupied molecular orbital (SOMO) indicates that the electron wavefunction inside the cavity and in the frontier orbitals of O atoms have opposite signs, suggesting a way to distinguish these two contributions. Plotting the isodensity contour maps of the SOMO for progressively increasing amplitudes helps to visualize these two contributions (as in Figure 1S(a), section 4.1). Typically, the diffuse, positive part of SOMO occupies 80-90% of the geometrical cavity at the density of $+(0.03-0.05)$ $a_0^{-3}$ and less than 10% at the density of $+(0.07-0.1)$ $a_0^{-3}$ (where $a_0 \approx 0.53$ Å is the atomic unit of length). In large clusters, ca. 50-60% of the total SOMO density is contained inside the sphere centered at $X$ corresponding to the closest of the NHB hydrogens (that is subsequently denoted as H$_a$); at least 75% of the total density is contained within the 3 Å sphere. The highest (negative) density is found in the frontal *O 2p* orbitals of oxygen atoms in the first solvation shell (Figures 1(a) and 1(b)).

Throughout the next section, little attention is paid to the energetics of the electron solvation. We are mainly interested in the structural aspects and the salient features of the ground state wavefunction. The radius of gyration $r_g$ for the electron given below is defined as $r_g = \left\langle r^2 - \langle r \rangle^2 \right\rangle$. We used SOMO for this averaging. In the one-electron model, the gyration radius can be roughly estimated from the optical spectrum moment analysis for the $s \rightarrow p$ absorption band, as described by Bartels; [58] the typical estimate is 2.5 Å. [58] The total spin density $\phi_{2p}^O$ in the *O 2p* orbitals of water (e.g., Figure 2(a)) was defined (consistently with the typical way in which such a parameter would be experimentally determined in EPR spectroscopy) as the sum of $\left| B_{zz}^O / B_{zz}^O(at.) \right|$ taken over all oxygen-17 nuclei.

## 4. Results and Discussion.

### 4.1. Small, symmetrical anions (n=4, 6, and 8).

Small water anion clusters observed experimentally in the gas phase either do not attach the electron or yield surface-bound electrons. [19] The resulting species are of great theoretical interest, but provide limited insight into the structure of trapped electron in liquids and glasses. Previous *ab initio* and DFT studies [14,15,20,21,22] suggest that internal solvation is possible when several (at least four) NHB hydrogens form a "solvation cavity." By contrast, H-bonded protons play almost no role in the electron solvation. The simplest anion cluster that has the desired properties is a tetrahedral (*D$_{2d}$* symmetrical) *b*-type cluster shown in Figure 1S (the Natori model of "solvated electron"). [14,15,25] This cluster (for the optimized geometry) has the lowest energy in the B3LYP/6-31+G** model, however, for larger basis sets (6-311++G** and aug-cc-VDZ) a *C$_{4h}$* symmetrical planar ring (Figure 1S(d)) has the lowest energy (the relative energies for these two clusters and *D$_{2d}$* symmetrical *d*-type cluster (Figure 1S(b)) and *C$_{4h}$* *b*-type planar cluster (Figure 1S(c)) are given in Table 1S). The energy switchover upon extension of the basis set follows the change from external to internal solvation. Despite considerable variation of the structure, all of these $b$- and $d$-type tetramer anions exhibit $a^0 \approx$ -(18-24) G and $a^H$ that is negative and small (for NHB hydroxyl groups). For $D_{2d}$ symmetrical $b$-type clusters, the comparison of hfcc parameters obtained using various methods is given in Table 2S. All of these methods yield a ground state that exhibits a diffuse positive density at the center (observe that the wavefunction is nonspherical) complemented by negative density in the frontal lobes of the $O\ 2p$ orbitals (Figure 1S(a)). Depending on the method and the basis set, the $r_{XH}$ distance (between the wavefunction centroid at X and the closest proton, $H_a$) varies between 1.45 Å (LSDA/aug-cc-pVDZ) and 3.2 Å (HF/aug-cc-pVDZ); for a given basis set, this distance is always longer for the HF method than for the B3LYP and MP2 methods (which yield similar optimized geometries). Since the size of the cluster largely defines the overlap of the SOMO wavefunction with the nuclei, comparing the hfcc parameters obtained for different optimized structures is not instructive. To facilitate such a comparison, we have calculated several parameters for $D_{2d}$ symmetrical clusters as a function of the X-$H_a$ distance $r_{XH,}$ optimizing all other degrees of freedom. The results are shown in Figures 2S to 5S; in Figure 6S these data are compared to the analogous results for a $d$-type tetrahedral anion.

As the cavity size increases from 1.8 to 3.2 Å, the total spin density $\phi_{2p}^O$ in the $O\ 2p$ orbitals decreases from 0.11 to 0.04 (Figure 2S(a)), and the Mulliken spin density on oxygen atoms decreases from –0.2 to –0.1 (B3LYP/6-311++G** model; Figure 2S(b)). The atomic spin density on the $H_a$ protons is negative, which immediately suggests that $a^H < 0$. The isotropic hfcc's $a^O$ on oxygen-17 and the protons decrease exponentially with $r_{XH}$, from –25 to –14 G and –0.8 to –0.2 G, respectively (Figure 3S(a)). Observe that $a^H$ is small and negative for all cavity sizes. The isotropic and anisotropic hfcc's for $H_a$ and $H_b$ protons plotted vs. the distance $r_{XH}$ to these nuclei follow the same general dependence (Figure 3S(a) and 3S(b)). Only for $r_{XH}$>3 Å does the constant $B_{zz}^H$ approach the estimate given by the point-dipole approximation, eq. (A9) (solid line in Figure 3S(b)); at shorter distances the anisotropic hfcc is significantly lower than this point-dipole estimate. The estimates for $B_{zz}^H$ (as is the case for all other water anions) obtained using B3LYP and HF methods are very close (Figure 4S(b)). For $r_{XH} \approx 2$ Å, $B_{zz}^H \approx 4$ G instead of 7.2 G in the point dipole approximation (Figures 3S(b) and 4S(b)). To obtain the experimental estimate of $\approx 7$ G, the X-$H_a$ distance should be $< 1.5$ Å, which is unrealistic. The experimental $a^H$ (-0.93 G) [32,33] can be matched only for $r_{XH} < 1.6$ Å (Figure 3S(a)). Thus, the tetrahedral arrangement seems to be excluded by our results. This is also suggested by Figure 5S *(to the bottom)* that shows the plot of the contribution to the second moment from the protons, $M_2^H$. For $r_{XH} \approx 1.8\text{-}2$ Å, this parameter is only 10-15 $G^2$, which is significantly smaller than experimental 21-23 $G^2$. [28,29,32] This is due to the smallness of the anisotropic contribution (Figure 3S(a)); the isotropic contribution to the EPR line width is negligible.

In the HF model (with the same basis set, Figure 4S(a)), the isotropic hfcc on the protons are several times more negative than in the B3LYP model (for $H_a$ changing from $-5.2$ G to $-1.4$ G when the X-$H_a$ distance changes from 1.8 to 3.2 Å). Such estimates are clearly incompatible with the experimental ones. For isotropic hfcc constants on oxygen-17, the HF methods always yield $a^O$ that are 20-100% less negative than B3LYP (Figure 4S(a)), resulting in smaller estimates for $M_2^O$ (which is dominated by these isotropic hfcc's. Figure 5S). Either way, the latter parameter is a few thousands of $G^2$, which is significantly greater than 360 $G^2$ given by Schlick et al. [30] (see section 2). All of these considerations suggest that tetrahedral sites for $e_{hyd}^-$ are rejected by EPR and ESEEM results. The same reasoning excludes *d*-type tetrahedral clusters (Figure 6S summarizes various parameters). Our conclusion is in full accord with MQC MD simulations indicating that the coordination number of the electron is close to six. [8-12] We have examined such clusters for two reasons. First, all large anion clusters known to trap the electron internally from previous *ab initio* and DFT studies have tetrahedral core anions. As shown below, having more water molecules does not qualitatively change the analysis given above. Second, it is clearly seen that the number of nearby water molecules has to be relatively large; thereby the analysis of Astashkin et al. [32] suggesting just two water molecules in the first solvation shell cannot be correct.

We turn to the octahedral complexes (Figure 1S(a) and Figures 2 through 4). Such complexes are expected to resemble most closely the "real" hydrated electron in liquid water. There are important differences between the results for *octa-* and *tetra-* hedral complexes. These differences are traceable to the greater sphericity of the electron wavefunction and more extensive spin sharing in the larger anions (as seen from Figure 2(a)). First, the $B_{zz}^H$ more closely follow the point-dipole model (Figure 3(b)); thus, it is possible to match the experimental estimate of this parameter for X-H$_a$ distance of 1.8-2 Å. Matching of the experimental $a^H$ [32] is possible for $r_{XH} \approx$ 2-2.1 Å (Figure 3(a)), and matching of the experimental $M_2^H$ [28,29,32] gives $r_{XH} \approx 2$ Å (Figure 4). Thus, all three EPR parameters for the protons can be simultaneously matched for the same cavity geometry. This matching is possible only in the DFT models: as was the case with the tetrahedral clusters, HF and MP2 methods grossly overestimate $a^H$ yielding, for realistic cavity sizes ($r_{XH} < 2.5$ Å), negative hfcc's of several Gauss, values that are excluded by ESEEM spectroscopy; [32,33] furthermore, large $a^H$ would increase $M_2^H$ to 50-80 G$^2$, which is inconsistent with EPR results. The estimate for $M_2^O$ is ca. 5000 G$^2$ (Figure 4, ca. 3000 G$^2$ in the HF model), which suggests a line width $\Delta B_{pp}$ of ca. 140 G (for fully oxygen-17 substituted $e_{hyd}^-$). The energy minimum is at $r_{XH} \approx 2.1$ Å in the B3LYP model (see the SOMO maps in Figure 1(a)) and ca. 3 Å in the HF model.

Figure 2(a) shows the cavity size dependence for the total population $\phi_{2p}^O$ of $O$ $2p$ orbitals and the gyration radius $r_g$. As $r_{XH}$ increases from 1.8 to 3.2 Å, the spin density transferred to oxygen atoms decreases from 0.20 to 0.05, and the gyration radius $r_g$ increases from 2.4 to 3.6 Å (B3LYP/6-311++G** model). Once more, the experimental $r_g \approx 2.5$ Å [58] is matched for $r_{XH} \approx 1.9$-2.0 Å (Figure 2(a)). Figure 2(b) shows the cavity size dependence for atomic spin and charge densities obtained by Mulliken population analysis. As the cavity increases, the charge on $H_a$ and $O$ in the hydroxyl groups gradually approaches its value for individual water molecules (in the same model), +0.25 and –0.5 $e$. At $r_{XH} \approx 2$ Å, the corresponding atomic charges are –0.05 and –0.31 $e$, and the atomic spin densities are +0.27 and –0.17, respectively (the spin density is always negative for H$_a$ protons).

Finally, we briefly consider the results obtained for two cube shaped $n$=8 water anions: a $D_4$ symmetrical $b$-type anion shown in Figure 1(b) and the $C_{4h}$ symmetrical $d$-type anion shown in Figure 7S (see Figure 8S for the summary of EPR parameters). Since the former anion has high degree of sphericity, the point approximation does not break down even for X-H$_a$ distances as short as 2 Å (Figure 8S(d)). The degree of spin transfer to *O 2p* orbitals is greater than in the octahedral and tetrahedral anions ($\phi^O_{2p} \approx 0.3$ for $r_{XH} \approx 1.8$ Å). The isotropic constants $a^H$ (Figure 8S(a-c)) are not too different from those for octahedral anions; since the coordination number is greater, the second moment is too large: $M^H_2 \approx 48$ G$^2$ (vs. experimental 21-23 G$^2$) [28,29,32] for $r_{XH} \approx 2$ Å. Even for the *d*-type anion (in which the electron is solvated by both OH groups of the water molecule, Figure 7S) the isotropic hfcc's on the protons are small and negative (ca. $-1$ G for $r_{XH} \approx$ 2-2.5 G, Figure 8S(b)). When $^{17}$O constants estimated for these anions are plotted against $r_{XO}$, the hfcc's for both types of clusters follow each other, suggesting that $a^O$ is mainly a function of the X-O distance rather than molecule orientation (Figure 8S(a)). From the standpoint of EPR parameters, the major difference between the *b*- and *d*-orientation appears to be in the anisotropic constants for the inner protons (Figure 8S(d)): while for the *b*-type anion the point approximation is accurate, for the *d*-type anion (and this refers to all anion clusters that we examined), $B^H_{zz}$ is significantly smaller than the estimate obtained in the point dipole approximation. Consequently, the estimates for this parameter become unrealistically small, and *d*-orientation of water molecules is not supported by our simulations.

All of these results provide strong support for "Kevan's" octahedral model [24,25] with the preferential orientation of H-O groups towards the center of the solvation cavity. This model, despite its being a gross idealization of $e^-_{hyd}$, appears to capture several important features that are observed for the hydrated electron. However, using this model for *quantitative* simulation of ESEEM spectra (rather than the extracted "mean" hfcc's) gives poor results. The structure of $e^-_{hyd}$ is neither regular nor octahedral; there is considerable variation in the coordination number, orientation of water molecules, etc. Many such conformations should be averaged to obtain the distributions of observable parameters. That is done in Part 2 of this study. [7]

### 4.2. Large anions (n=20 and n=24)

An *n*=20 cluster (***w20n2*** anion found by Khan [20]) and three *n*=24 clusters (***t24n1*** cluster found by Khan [20] and ***4⁶6⁸B*** and ***5¹²6²B*** anions found by Herbert and Head-Gordon) [21] were examined using B3LYP and HF methods. The electron in these clusters is bound internally by four NHB hydroxyl groups (the four SOMO maps are shown in Figures 9S to 12S); for $4^66^8B$ anion these OH groups are arranged in a rectangle, for the other three clusters the arrangement is tetrahedral. The mean distance $\langle r_{XH} \rangle$ to the nearest hydroxyl protons is 1.78, 1.87, 2.21, and 2.16 Å, respectively (Table 3S). Remarkably, for these large cluster anions, even relatively tight binding basis sets (such as 6-31+G**) give estimates for *average* $^{17}O$ and $^1H$ hfcc constants that are comparable to the averages obtained using more diffuse basis sets (Table 4S). Isotropic hfcc's for protons are small ($|a^H| < 1$ G): either slightly negative or slightly positive (Table 3S). That relates only to DFT calculations (Table 3S); with the HF method, as was the case for smaller water anions, one obtains large negative $a^H$ of –(5-8) G (Table 4S). Consequently, the estimates for $M_2^H$ are > 100 $G^2$ for Khan's [20] clusters (5 times greater than the experimental estimates). On the other hand, the estimates for $M_2^H$ obtained using B3LYP method for $4^66^8B$ and $5^{12}6^2B$ anions are < 10 $G^2$, which is unrealistically small. This is due to the large cavity size (as compared to Khan's clusters) and small coordination number. Only for $w20n2$ and $t24n1$ anions (for which $\langle r_{XH} \rangle < 1.85$ Å), is $B_{zz}^H$ for the nearest protons close to the experimental value; for larger clusters ($r_{XH} > 2.16$ Å), $B_{zz}^H < 4$ G. For other than the nearest four protons, the isotropic hfcc's are very small ($|a^H| < 0.05$ G for B3LYP method; $a^H < 0.2$ G for the HF method, Tables 3S and 4S), i.e., the excess electron is localized in the first solvation shell. The isotropic hfcc's for oxygen-17 nuclei suggest the same: for oxygen-17 nuclei in the first solvation shell, the mean $a^O$ ranges from –24.5 G to –17.2 G (decreasing in absolute value for larger cavities), whereas for the second solvation shell these constants range from –3 to –4 G (Table 3S). The total population of *O 2p* orbitals, despite this partial spin transfer to the second shell oxygen atoms, is quite small (as was the case for tetrahedral cavities examined in section 4.1), ca. 0.1-0.14 (Table 3S). Atomic spin densities for these oxygens are also small ((1-5)x10$^{-3}$ vs. –0.05 for NHB hydroxyl groups, Table 3S). Thus, the degree of electron density penetration beyond the first solvation shell is too small to have a significant effect on the second moment $M_2^O$ from oxygen-17, that is similar to those for isolated tetrahedral clusters with the same $r_{XH}$ (compare Table 3S and Figure 5S).

The comparison of EPR parameters for these large cluster anions with smaller tetrahedral anions obtained by retaining only the four water molecules forming the "solvation cavity" suggests that the effect of the second solvation shell on these EPR parameters is quite small. The second solvation shell is important for maintaining the fortuitous orientation of water molecules and obtaining the correct energetics; the EPR parameters, by contrast, depend mainly on the interaction of the excess electron density with *the nuclei in the first solvation shell.* This relatively tight localization of the SOMO justifies the use of the embedded cluster approach suggested in *Part 2* of this series. [7]

## 5. Conclusion.

This study aims to explain EPR/ESEEM parameters observed for trapped electrons in low-temperature alkaline ices. General considerations and experimental data (section 2 and Appendix A) suggest that such an explanation cannot be sought using familiar one-electron models of electron solvation. Hence *ab initio* and DFT methods were used to calculate hyperfine coupling tensors for water anion clusters that internally localize the electron via interaction with 4-8 NHB hydroxyl groups. Both small ($n$=4-8) and large ($n$=20,24) model cluster anions were examined. For small clusters, the effects of coordination number and cavity size were studied. The comparison of small tetrahedral clusters with larger clusters identified by Khan [20] and Herbert and Head-Gordon, [21] in which the electron is 4-coordinated suggests that the electron wave function is localized mainly over this first solvation shell and thus these small clusters are representative of the ones in the bulk water (this line of reasoning is continued and extended in Part 2). Examination of these small clusters suggests that 10-20% of the spin density is transferred into the frontal *O 2p* orbitals of the hydroxyl groups forming the cavity. This transfer has several consequences for the hfcc parameters. First, as a result of spin bond polarization, it makes isotropic hfcc's on the protons small and negative, in agreement with the ESEEM data of Astashkin et al. [32] Second, for clusters with low coordination number, it introduces significant lowering of anisotropic hfcc's as compared to point-dipole approximation, as was anticipated by Golden and Tuttle. [41] As the cavity sizes were determined using the latter approximation, [24,29,32,33] our results indicate the limited import of such estimates. Third, there is a very significant spin density on oxygen, suggesting that EPR results of Schlick et al. [30] for $^{17}O$ substituted glasses (used to justify the one-electron models) were compromised, as was also suggested by subsequent studies by the same authors (section 2). [49]

While the one-electron point-dipole octahedral ("Kevan's") model [24,25] might be overly simplistic, it turns out that such $b$-oriented octahedral arrangement of water molecules does capture several experimentally observed features of $e_{hyd}^-$; ironically, that occurs *only* in the multielectron model of the core anion. The isotropic and anisotropic hfcc tensor parameters determined from ESEEM spectra, [32,33] the second moments of the EPR spectra, [28,29,32] the downshifts of the H-O-H bend and H-O stretch modes, [6] and the gyration radius of the electron [58] – all these parameters can all be quantitatively accounted for in such an octahedral model for $r_{XH} \approx 2$-2.2 Å (see Part 2 [7] for vibrational analysis). The DFT model thus provides rationalization for all of the experimental observables involving the *ground state* wave function of $e_{hyd}^-$. None of these have been accounted for using the existing one-electron models.

Since MQC MD calculations [9-12] indicate that the coordination number of hydrated electron is ca. 6, our results suggest that the octahedral model is correct "on average." While $e_{hyd}^-$ does not have a regular solvation shell, like the idealized cluster anions examined in section 4, this average does not look too dissimilar to the octahedral model, if one looks at mean parameters. This is demonstrated in Part 2 of this study. [7] While the magnetic parameters for different trapping sites show considerable variation, the mean values are similar to the ones obtained in simple cluster models provided that the mean cavity size is the same. This is, again, due to the highly localized nature of the ground state electron wavefunction.

In many ways, the picture of the excess electron in water that emerges from the DFT model is similar to the familiar one-electron picture of this species: a large fraction of the excess electron density (ca. 50-60% of the SOMO) is contained inside the cavity; NHB hydroxyl groups stabilize the electron, there is little spin density in the hydroxyl protons, and the electron wavefunction in the cavity has $s$-character. Yet it also departs from this picture. A substantial fraction (10-20%) of the spin density is in the oxygen atoms of these NHB groups, so $e_{hyd}^-$ can be viewed as a multimer radical anion. [1,2,16] That fraction is smaller for $e_{hyd}^-$ than for the ammoniated electron, [1] where most of the spin density is in the *N 2p* orbitals: the hydrated electron is, perhaps, the closest one can get to the one-electron picture, hence the remarkable success of the latter in rationalizing the experimental observations.

## 6. Acknowledgement.

I thank B. J. Schwartz, S. E. Bradforth, R. Larsen, D. M. Bartels, J. M. Herbert, A. Khan, and W. Domcke for many useful discussions and M. C. Sauer, Jr. for careful reading of the manuscript. This work was supported by the Office of Science, Division of Chemical Sciences, US-DOE under contract number W-31-109-ENG-38.

***Supporting Information Available:*** A single PDF file containing (1) Appendix A: EPR and ESEEM spectra of trapped electrons: the basics; (2.) Appendix B: Geometries of large anion clusters; (3.) Tables 1S to 4S; (4.) Figures 1S to 12S with captions. This material is available free of charge via the Internet at http://pubs.acs.org.

**References.**

(1)     Shkrob I. A. *J. Phys. Chem. A* **2006**, *110*, 3967.

(2)     Shkrob, I. A.; Sauer, Jr., M. C. *J. Phys. Chem. A* **2006**; *in press*; preprint available
        on http://arxiv.org/abs/physics/0512142

(3)     Shkrob, I. A.; Sauer, Jr., M. C. *J. Phys. Chem. A* **2005**, *109*, 5754; *J. Chem. Phys.*
        **2005**, *122*, 134503.

(4)     Shkrob, I. A.; Sauer, Jr., M. C. *J. Phys. Chem. A* **2002,** *106*, 9120; *see also*
        Shkrob, I. A.; Takeda, K.; Williams, F. *J. Phys. Chem. A* **2002,** *106*, 9132; Xia,
        C.; Peon, J.; Kohler, B. *J. Chem. Phys.* **2002**, *117*, 8855.

(5)     Hart, E. J.; Anbar, M. *The Hydrated Electron*, Wiley-Interscience: New York,
        1970.

(6)     (a) Tauber, M. J.; Mathies, R. A. *Chem. Phys. Lett.* **2002**, *354*, 518 and(b) *J. Phys.
        Chem. A* **2001**, *105*, 10952; Tauber, M. J.; Mathies, R. A. *J. Am. Chem. Soc.*
        **2003**, *125*, 1394; (b) Mizuno, M.; Tahara, T. *J. Phys. Chem. A* **2001**, *105*, 8823
        and *J. Phys. Chem. A* **2003**, *107*, 2411; Mizuno, M.; Yamaguchi, S.; Tahara, T. In
        *Femtochemistry and Femtobiology*, Martin, M. M.; Hynes, J. T., Eds.; Elsevier,
        Amsterdam, The Netherlands, 2004; pp. 225.

(7)     Shkrob, I. A.; Larsen, R.; Schwartz, B. J. *J. Phys. Chem. A* **2006**, Part 2 of this
        series.

(8)     Schnitker, J.; Rossky, P. J. *J. Chem. Phys.* **1986**, *86*, 3471; Wallqvist, A.;
        Thirumalai, D.; Berne, B. J. *J. Chem. Phys.* **1986**, *86*, 6404; Romero, C.; Jonah,
        C. D. *J. Chem. Phys.* **1988**, *90*, 1877; Wallqvist, A.; Martyna, G.; Berne, B. J. *J.
        Phys. Chem.* **1988**, *92*, 1721; Miura, S.; Hirata, F. *J. Phys. Chem.* **1994**, *98*, 9649.

(9)     Schnitker, J.; Rossky, P. J.; Kenney-Wallace, G. A. *J. Chem. Phys.* **1986**, *85*, 2989;
        Rossky, P. J.; Schnitker, J. *J. Phys. Chem.* **1988**, *92*, 4277; Schnitker, J.;
        Motakabbir, K.; Rossky, P. J.; Friesner, R. A. *Phys. Rev. Lett.* **1988**, *60*, 456;

Webster, F. J.; Schnitker, J.; Frierichs, M. S.; Friesner, R. A.; Rossky, P. J. *Phys. Rev. Lett.* **1991**, *66*, 3172; Webster, F. J.; Rossky, P. J.; Friesner, R. A. *Comp. Phys. Comm.* **1991**, *63*, 494; Motakabbir, K.; Schnitker, J.; Rossky, P. J. *J. Chem. Phys.* **1992**, *97*, 2055; Murphrey, T. H.; Rossky, P. J. *J. Chem. Phys.* **1993**, *99*, 515.

(10)  Schwartz, B. J.; Rossky, P. J. *J. Chem. Phys.* **1994**, *101*, 6917; *J. Phys. Chem.* **1994**, *98*, 4489; *Phys. Rev. Lett.* **1994**, *72*, 3282; *J. Chem. Phys.* **1994**, *101*, 6902; Rosenthal, S. J.; Schwartz, B. J.; Rossky, P. J. *Chem. Phys. Lett.* **1994**, *229*, 443.

(11)  Wong, K. F.; Rossky, P. J. *J. Phys. Chem. A* **2001**, *105*, 2546.

(12)  Borgis, D.; Staib, A. *Chem. Phys. Lett.* **1994**, *230*, 405; Staib, A.; Borgis, D. *J. Chem. Phys.* **1995**, *103*, 2642; Borgis, D.; Staib, A. *J. Chim. Phys.* **1996**, *39*, 1628; *J. Chem. Phys.* **1996**, *104*, 4776; *J. Phys.: Condens. Matter* **1996**, *8*, 9389; Staib, A.; Borgis, D. *J. Chem. Phys.* **1996**, *104*, 9027; Borgis, D.; Bratos, S. *J. Mol. Struct.* **1997**, *1997*, 537; Nicolas, C.; Boutin, A.; Levy, B.; Borgis, D. *J. Chem. Phys.* **2003**, *118*, 9689.

(13)  Ogg, R. A. *J. Am. Chem. Soc.* **1940**, *68*, 155; *J. Chem. Phys.* **1946**, 14, 114 and 295; *Phys. Rev.* **1946**, *69*, 243 and 668; Jortner, J. *J. Chem. Phys.* **1959**, *30*, 839; Kestner, N. R. In *Electrons in Fluids*; Jortner, J., Kestner, N. R., Eds.; Springer-Verlag: New York, 1973; pp 1.; (17) Sprik, M.; Impey, R. W.; Klein, M. L. *J. Chem. Phys.* **1985**, *83*, 5802; Sprik, M.; Impey, R. W.; Klein, M. L. *Phys. Rev. Lett.* **1986**, *56*, 2326; Sprik, M.; Klein, M. L. *J. Chem. Phys.* **1989**, *91*, 5665; *J. Chem. Phys.* **1988**, *89*, 1592; Marchi, M.; Sprik, M.; Klein, M. L. *J. Phys. Chem.* **1990**, *94*, 431; Rodriguez, J.; Skaf, M. S.; Laria, D. *J. Chem. Phys.* **2003**, *119*, 6044.

(14)  Newton, M. D. *J. Phys. Chem.* **1975**, *79*, 2795.

(15)  Clark, T.; Illing, G. *J. Am. Chem. Soc.* **1987**, *109*, 1013.

(16)  Symons, M. C. R. *Chem. Soc. Rev.* **1976**, *5*, 337.

(17)     Thompson, J. C. *Electrons in Liquid Ammonia*; Clarendon Press: Oxford, 1976.

(18)     Nielsen, S. O. ; Michael, B. D.; Hart, E. J. *J. Phys. Chem.* **1976**, *80*, 2482.

(19)     Ayotte, P.; Johnson, M. A. *J. Chem. Phys.* **1997**, *106*, 811 and *J. Chem. Phys.* **1999**, *110*, 6268; Weber, J. M. et al. *Chem. Phys. Lett.* **2001**, *339*, 337; Coe, J. V.; Lee, G. H.; Eaton, J. G.; Arnold, S.; H.W. Sarkas; Bowen, K. H.; Ludewigt, C.; Haberland, H.; Worsnop, D. *J. Chem. Phys.* **1990**, *92*, 3980; Coe, J. V.; Earhart, A. D.; Cohen, M. H.; Hoffman, G.J.; Sarkas, H.W.; Bowen, K.H. *J. Chem. Phys.* **1997**, **107**, 6023; Verlet, J. R. R.; Bragg, A. E.; Kammrath, A.; Chesnovsky, O.; Neumark, D. M. *Science* **2005**, *307*, 93 and *Science* **2004**, *306*, 669; Paik, D. H.; Lee, I-R.; Yang, D.-S.; Baskin, J. C.; Zewail, A. H. *Science* **2004**, *306*, 672; Headrick; Diken, E. G.; Roscioli, J. R.; Weddle, G. H.; Johnson, M. A. *Science* **2004**, *306*, 675; Hammer, N. I.; Roscioli, J. R.; Johnson, M. A. *J. Phys. Chem. A* **2005**, *109*, 7896 and references therein.

(20)     Khan, A. *Chem. Phys. Lett*. **2005**, *401*, 85; *J. Chem. Phys.* **2003**, *118*, 1684; *J. Chem. Phys.* **2003**, *121*, 280.

(21)     Herbert, J. M.; Head-Gordon, M. *J. Phys. Chem. A* **2005**, *109*, 5217; *Phys. Chem. Chem. Phys.* **2006**, *8*, 68.

(22)     Kim, K. S.; Park, I.; Lee, K.; Cho, K.; Lee, J. Y.; Kim, J.; Joannopoulos, J. D. *Phys. Rev. Lett.* **1996**, 76, 956; Kim, K. S.; Lee, S.; Kim, J.; Lee, J. Y. *J. Am. Chem. Soc.* **1997**, *119*, 9329; Lee H. M.; Kim, K. S. *J. Chem. Phys.* **2002**, *117*, 706; Lee, H. M.; Suh, S. B.; Kim, K. S. *J. Chem. Phys.* **2003**, *118*, 9981; *J. Chem. Phys.* **2003**, *119*, 187.

(23)     Boero, M.; Parrinello, M.; Terakura, K.; Ikeshoji, T.; Liew, C. C. *Phys. Rev. Lett.* **2003**, *90*, 226403.

(24)     Narayana, P. A.; Bowman, M. K.; Kevan, L.; Yudanov, V. F.; Tsvetkov, Yu. D. *J. Chem. Phys.* **1975**, *63*, 3365.

(25)    (a) Natori, M.; Watanabe, T. *J. Phys. Soc. Jap.* **1966**, *21*, 1573; (b) Natori, M. *J. Phys. Soc. Jap.* **1968**, *24*, 913, ibid. **1969**, *27*, 1309.

(26)    Helbert, J.; Kevan, L.; Bales, B. L. *J. Chem. Phys.* **1972**, *57*, 723.

(27)    Yoshida H.; Feng. D.-F.; Kevan, L. *J. Chem. Phys.* **1973**, *58*, 3411.

(28)    Bales, B. L.; Helbert, J.; Kevan, L. *J. Phys. Chem.* **1974**, *78*, 221.

(29)    Bales B. L.; Bowman, M. K.; Kevan, L.; Schwartz, R. N. *J. Chem. Phys.* **1975**, *63*, 3008.

(30)    Schlick, S.; Narayana, P. A.; Kevan, L. *J. Chem. Phys.* **1976**, *64*, 3153.

(31)    Dikanov S. A.; Tsvetkov, Yu. D.; Astashkin, A. V.; Bowman, M. K. *Chem. Phys. Lett.* **1983**, *94*, 231.

(32)    Astashkin, A. V.; Dikanov, S. A.; Tsvetkov, Yu. D. *Chem. Phys. Lett.* **1988**, *144*, 258.

(33)    Dikanov S. A.; Tsvetkov, Yu. D. "*Electron Spin Echo Envelope Modulation (ESEEM) Spectroscopy*," CRC Press: Boca Raton, 1992; Ch. 133.II, pp. 244-251.

(34)    (a) Bednarek, J.; Plonka, A.; Hallbrucker, A.; Mayer, E.; Symons, M. C. R. *J. Am. Chem. Soc.* **1996**, *118*, 9378; *Radiat. Phys. Chem.* **1998**, *53*, 635; *J. Phys. Chem. A* **1998**, *102*, 9091; *Phys. Chem. Chem. Phys.* **2000**, *2*, 1587; (b) Gillis, H. A.; Quickenden, T. I. *Can. J. Chem.* **2001**, *79*, 80 *and references therein*; (c) Johnson, J. E.; Moulton, G. C. *J. Chem. Phys.* **1978**, *69*, 3108; (d) Hase, H.; Kawabata, K. *J. Chem. Phys.* **1976**, *65*, 64.

(35)    de Haas, M. P.; Kunst, M.; Warman, J.; Verberne, J. B. *J. Phys. Chem.* **1983**, *87*, 4089 and 4096.

(36)    Nilsson, G. *J. Chem. Phys.* **1972**, *56*, 3437.

(37)    Bennett, J. E.; Mile, B.; Thomas, A. *J. Chem. Soc.* **1967A**, 1393; 1969, 1502.

(38)   Ershov, B. G.; Pikaev, A. K. *Radiation Effects* **1969**, *2*, 135.

(39)   Ohno, K.; Takemura, I.; Sohma, J. Chem. Phys. **1972**, 56, 1202.

(40)   Atkins, P. W.; Symons, M. C. R. *The Structure of Inorganic Radicals*, Elsevier: Amsterdam, 1967.

(41)   Golden S.; Tuttle, Jr., T. R. *J. Phys. Chem.* **1984**, *88*, 3781.

(42)   Kroh, J.; Noda, S.; Yoshida, K.; Yoshida, H. *Bull. Chem. Soc. Jpn.* **1978**, *51*, 1961; Zagorski, Z. P.; Grodkowski, J.; Bobrowski, K. *Radiat. Phys. Chem.* **1980**, *15*, 343; *Chem. Phys. Lett.* **1978**, *59*, 533; Kondo, Y.; Aikawa, M.; Sumioshi, T.; Katayama, M.; Kroh, J. *J. Phys. Chem.* **1980**, *84*, 2544; Wolszczk, Wypych, M.; Tomczyk, M.; Kroh, J. *Res. Chem. Intermed.* **2002**, *28*, 537.

(43)   Spezia, R.; Nicolas, C.; Archirel, P.; Boutin, A. *J. Chem. Phys.* **2004**, *120*, 5261.

(44)   Fueki, K.; Feng. D.-F.; Kevan, L. *J. Phys. Chem.* **1970**, *74*, 1977; Fueki, K.; Feng. D.-F.; Kevan, L.; Christoffersen, R. E. *J. Phys. Chem.* **1971**, *75*, 2297; Fueki, K.; Feng. D.-F.; Kevan, L. *J. Am. Chem. Soc.* **1973**, *95*, 1398; Feng. D.-F.; Ebbing, D.; Kevan, L. *J. Chem.* Phys. **1974**, *61*, 249.

(45)   Noell, J. O.; Morokuma, K. *Chem. Phys. Lett.* **1975**, *36*, 465.

(46)   Schwartz, R. N.; Bowman, M. K.; Kevan, L. *J. Chem. Phys.* **1974**, *60*, 1690; Bales, B. L.; Kevan, L. *J. Chem. Phys.* **1974**, *60*, 710; Narayana, M.; Kevan, L. J. Chem. Phys. **1980**, *72*, 2891.

(47)   Narayana, M.; Kevan, L.; Samskog, P. O.; Lund, A.; Kispert, L. D. *J. Chem. Phys.* **1984**, *81*, 2297; Hase, H.; Ngo, F. Q. H.; Kevan, L. *J. Chem. Phys.* **1975**, *62*, 985.

(48)   Lund, A.; Schlick, S. *Res. Chem. Intermed.* **1989**, *11*, 37; Box, H. C.; Budzinski, E. E.; Freund, H. G.; Potter, W. R. *J. Chem. Phys.* **1979**, *70*, 1320 and 5040.

(49)   Schlick, S.; Kevan, L. *J. Phys. Chem.* **1977**, *81*, 1083.

(50)     Symons, M. C. R. *J. Chem. Soc. Farad. Trans. 1*, **1982**, *78*, 1953.

(51)     Becke, A. D. *Phys. Rev. A* **1988**, *38*, 3098

(52)     Lee, C., Yang, W.; Parr, R. G. *Phys. Rev. B* **1988**, *37*, 785

(53)     Frisch, M. J. et al, *Gaussian 98, revision A.1*, Gaussian, Inc., Pittsburgh, Pennsylvania, 1998.

(54)     MØller, C.; Plesset, M. C. *Phys. Rev.* **1934**, *46*, 618; Frish, M. J.; Head-Gordon, M.; Pople, J. A. *Chem. Phys. Lett.* **1988**, *153*, 503, ibid. **1990**, *166*, 275; ibid. **1990**, *166*, 281.

(55)     Wilson, A.; van Mourik, T.; Dunning, Jr., T. H. *J. Mol. Struct. (Theochem)* **1997**, *388*, 339 and references therein.

(56)     Barone, V. In *Recent Advances in DFT methods, Part I,* Ed. Chong, D. P., Word Scientific: Singapore, 1996.

(57)     Bradforth, S. E.; Jungwirth, P. *J. Phys. Chem. A* **2002**, *106*, 1286.

(58)     Bartels, D. M. *J. Chem. Phys.* **2001**, *115*, 4404.

**Figure Captions**

**Figure 1**

Isodensity maps for singly occupied molecular orbit (SOMO) of (a) $C_i$ symmetrical octahedral and (b) $D_4$ symmetrical cube shaped $b$-type water anions ($\pm 0.03$ $a_0^{-3}$ surfaces are shown, red is for positive, purple is for negative). The positive part of the wavefunction occupies the cavity; the negative part is shared between the O 2p orbitals of 6 or 8 water molecules forming the cavity. B3LYP/6-311++G** model for X-$H_a$ distance of 2.1 Å (optimized geometry).

**Figure 2**

(a) Total population $\phi_{2p}^O$ of oxygen $2p$ orbitals (filled circles, to the left) and radius of gyration, $r_g$ (open squares, to the right), as a function of $r_{XH}$, the X-$H_a$ distance (B3LYP/6-311++G** model for the octahedral water anion shown in Figure 1(a)). (b) Mulliken population analysis: atomic charge (filled symbols, top) and spin (empty symbols, bottom) densities for $H_a$ (circles), $H_b$ (triangles) and O (squares) atoms.

**Figure 3**

(a) Isotropic hfcc's for $^{17}$O (filled squares; to the top) and $^1$H (bottom) nuclei (the $H_a$ [filled circles] and $H_b$ [empty triangles] nuclei are shown together) vs. X-O and X-$H_{a,b}$ distances, respectively. Solid lines are exponential fits (B3LYP/6-311++G** calculation for the octahedral water anion); (b) The same as (a), for $zz$ principal component of anisotropic hfcc tensor, $B_{zz}^H$. The solid line is the estimate obtained in point-dipole approximation (eq. (A9)).

**Figure 4**

Second moments of EPR spectra for the octahedral water anion (B3LYP/6-311++G** model) vs. the cavity size (the X-$H_a$ distance). (a) The contribution from the $^{17}$O nuclei, (filled squares, top) (b) the proton contribution (filled circles, bottom); in the latter the (small) contribution from isotropic hfcc is shown by empty circles.

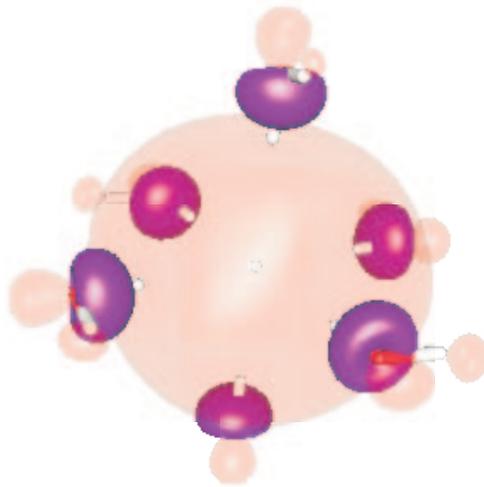

**(a)**

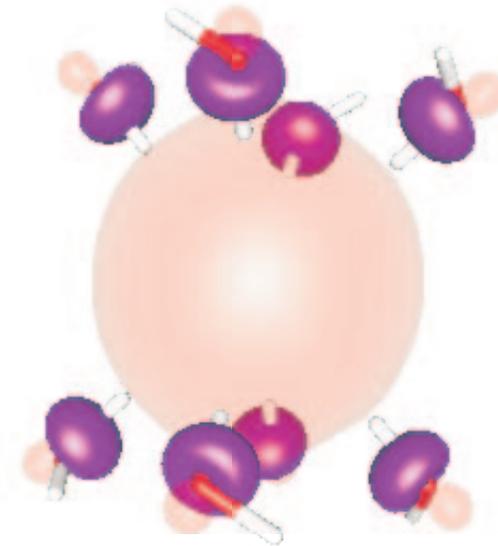

**(b)**

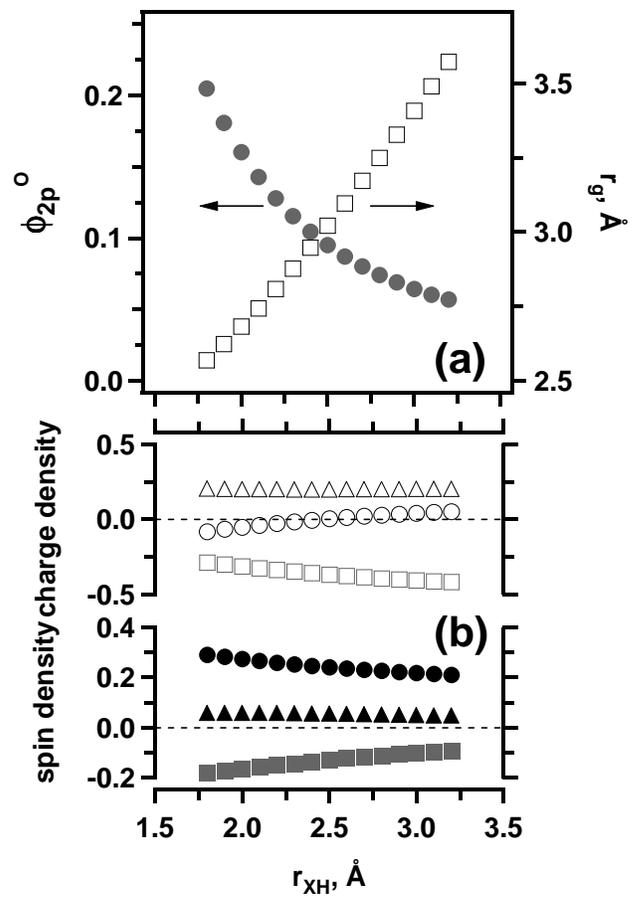

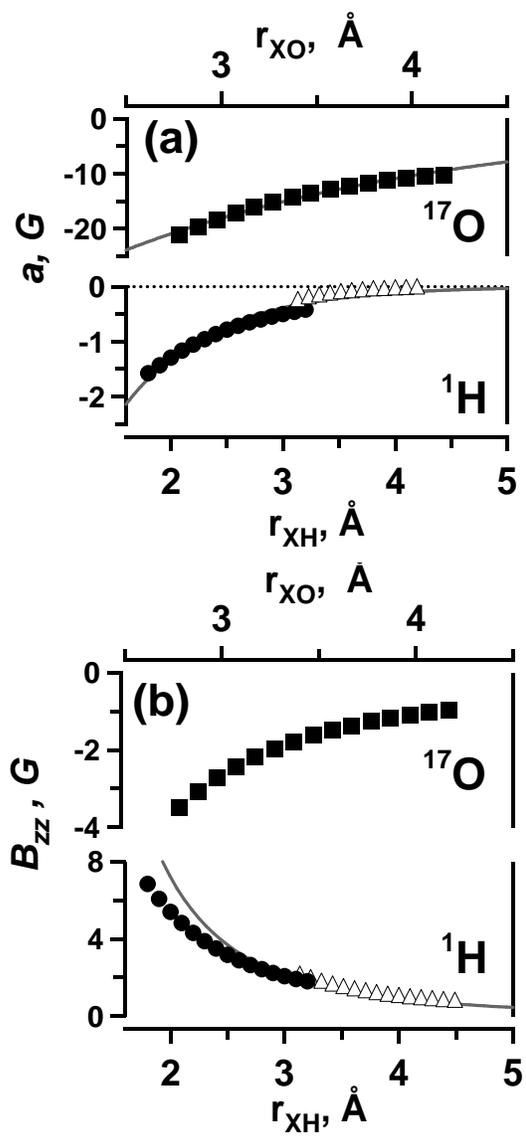

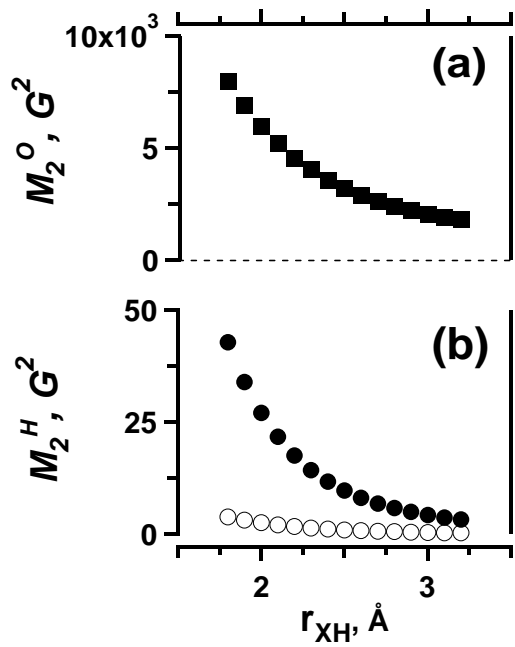

## Supporting Information.

**Appendix A. EPR and ESEEM spectra of trapped electrons: the primer.**

### 1. Electron paramagnetic resonance (EPR) spectra.

The spin Hamiltonian of a trapped electron is given by [40]

$$H = g\beta_e B_0 S_z - \sum_k \nu_k I_{z,k} + \sum_k S \bullet \mathbf{A}_k \bullet I_k \ , \tag{A1}$$

where $g$ is the g-factor of trapped electron (whose $g$-tensor is assumed to be fully isotropic), $\beta_e$ is Bohr magneton, $B_0$ is the magnetic field of the EPR spectrometer in the direction of axis $z$ of the laboratory frame, $S$ is the electron spin and $I_k$ is the nuclear spin of $k$-th nucleus, $\nu_k$ is the corresponding NMR frequency in the field of the EPR spectrometer, and $\mathbf{A}_k$ is the hyperfine coupling tensor with the principal values of $(A_{xx}, A_{yy}, A_{zz}) = (a + B_{xx}, a + B_{yy}, a + B_{zz})$ in the principal axis frame (which is different for every nucleus); $\mathbf{B}$ is the traceless tensor of anisotropic hyperfine interaction. The hfcc couplings are typically given in the field units (i.e., as $A_k/g\beta_e$). In eq. (A1), we have neglected nuclear quadrupole interaction for nuclei with $S > \frac{1}{2}$. The isotropic hyperfine coupling constant $a = (A_{xx} + A_{yy} + A_{zz})/3$ can be either positive or negative, even when the magnetic moment of the nucleus is positive. In the absence of spin polarization induced by the unpaired electron on a neighboring atom, the contact (Fermi) contribution

$$\frac{a}{g\beta_e} = \frac{8\pi}{3} g_n \beta_N \rho_s(0)$$

from the atomic $s$-function to isotropic hfcc is positive for $g_n > 0$ (where $\beta_N$ is the nuclear magneton and $g_n$ is the $g$-factor for the nucleus) because the spin density $\rho_s(0) = |\Phi_s(0)|^2$ on this nucleus is always positive. The negative contribution for protons in NHB hydroxyl groups (for which $g_n > 0$) originates through a spin bond polarization mechanism (that can be regarded as a kind of configuration interaction, see section II.5 in ref. 40) that involves the $sp^2$ electron in the oxygen atoms: in accordance with the Pauli principle, the spin of the electron in the $s$-orbital of the H atom has preferential antiparallel orientation with regard to the unpaired electron in the $p$-orbital of the oxygen atom. This imbalance results in negative spin density $\rho_s^H$ for the hydrogen atom, which in turn accounts for $a < 0$ (see, for example, Figure 2 and Table 3S that give atomic spin densities for $H_a$ atoms).

Anisotropic part of the hfcc tensor is given by

$$B_{ik} = g\beta g_n \beta_N \left\langle r^{-5}\left(3r_i r_k - r^2 \delta_{ik}\right)\right\rangle_\rho$$

where $i, k = x, y, z$ and $r_i$ are the components of the radius vector connecting the electron and the magnetic nucleus; the averaging is over the unpaired electron density.

We will assume that the tensor $\mathbf{A}$ is approximately axial, so

$$A_{xx,yy} \approx a + T_\perp \text{ and } A_{zz} \approx a - 2T_\perp \ , \tag{A2}$$

where $a$ is the isotropic hfcc, and $T_\perp$ is the perpendicular component of the anisotropic hfcc tensor $\left(T_\perp, T_\perp, -2T_\perp\right)$ (following the conventions of Chapter 6.I, ref. 33). For the electron in a lone $2p$ orbital, the $A$ tensor is axial and $T_\perp = -\frac{2}{5} g\beta g_n \beta_N \left\langle r^{-3}\right\rangle_{np}$ .

The second term in eq. (A1) is small and it matters only for the ESEEM experiment examined below. In a sufficiently large field, using perturbation theory the resonance field $B_{res}$ for the given frequency $\omega$ of the EPR spectrometer and the set of spin orientations $m_{z,k} = \{-I_k,...,+I_k\}$ (of the $I_{z,k}$ component of nuclear spin on the z axis of the laboratory frame) is given by

$$\omega = g\beta_e B_{res} + \sum_k \left|\mathbf{u} \bullet \mathbf{A}_k\right| \ m_{z,k} \ , \tag{A3}$$

where $\mathbf{u}$ is the unit vector in the direction of the magnetic field. For every set of the projections $\left\{m_{z,k}\right\}$ and the given set $\left\{\mathbf{A}_k\right\}$ one obtains an offset resonance field $\Delta B = B_{res} - \omega/g\beta_e$ ; the histogram of such fields gives the EPR spectrum in the absence of homo- and heterogeneous broadening. The latter can be taken into account by convoluting the histogram with a suitable function, typically a Gaussian one. The EPR spectra $G(\Delta B)$ are commonly presented as the first derivatives $G'(\Delta B)$ of such convoluted spectra with respect to the field (since the EPR spectra are often obtained by modulation of the external field). Observe that the spectrum is symmetrical with respect to the field around the origin at $\Delta B = 0$ . When the number of magnetic nuclei is large, the histogram and the spectrum become Gaussian,

$$G(\Delta B) \propto \exp\left[-2\left(\Delta B/\Delta B_{pp}\right)^2\right] \ , \tag{A4}$$

where $\Delta B_{pp}$ is the peak-to-peak width for $G'(\Delta B)$ . The most convenient way of simulating the EPR spectra using eq. (A3) is a Monte-Carlo method, in which vectors $\mathbf{u}$ and sets of projections $\left\{m_{z,k}\right\}$ are randomly generated; typically a sample of $10^6$ such sets yields a good quality EPR spectrum. The Monte-Carlo method also makes it very easy to take into account different isotopic configurations, by giving different weights to different sets of $\left\{\mathbf{A}_k\right\}$ of hfcc tensors.

The second moment $M_2$ of the EPR spectrum is defined as

$$M_2 = \int\limits_{-\infty}^{+\infty} \Delta B^2 G(\Delta B)\, d\Delta B \left/ \int\limits_{-\infty}^{+\infty} G(\Delta B)\, d\Delta B \right. ,  \qquad (A5)$$

which is, for a Gaussian line, given by

$$M_2 \approx \Delta B_{pp}^2 / 4  \qquad (A6)$$

When the number of magnetic nuclei is large, the hfcc constants are comparable to each other, and the EPR spectrum is Gaussian, the second moment can be estimated as

$$M_2 \approx \sum_k \left\langle \mathbf{A}_k^2 m_{z,k}^2 \right\rangle = \frac{1}{3} \sum_k I_k (I_k + 1) \left\langle A_k^2 \right\rangle ,  \qquad (A7)$$

where

$$\left\langle A_k^2 \right\rangle = 1/3 (\mathbf{A}_k : \mathbf{A}_k) = a + 1/3 \left( B_{xx,k}^2 + B_{yy,k}^2 + B_{zz,k}^2 \right) \approx a + 2T_\perp^2 .  \qquad (A8)$$

As seen from eq. (A7), the contributions from $^1$H and $^{17}$O nuclei to the second moment can be calculated separately (designated as $M_2^H$ and $M_2^O$); the total second moment $M_2 = M_2^H + \xi^O M_2^O$, where $\xi^O$ is the atomic fraction of $^{17}O$ nuclei (that can be equated to zero in the absence of isotope enrichment). These two contributions can be further divided into those from the isotropic and anisotropic parts of the corresponding hfcc tensors (eq. (A8)). The first contribution dominates for $^{17}$O, the second – for the protons. Figure A1 illustrates the formation of a Gaussian spectrum for a multinuclear system (a $n$=20 water anion, **w20n2**, [20] in B3LYP/6-311++G** model). The EPR spectrum for a fixed geometry cluster is angle averaged. Despite this fixed geometry, the EPR spectrum is nearly Gaussian due to the large number of protons coupled (Figure A1, part (a)); some structure can still be seen in the first derivative of this spectrum. For the $^{17}$O substituted cluster (37 at% $^{17}$O; Figure A1, part (b)), where the four strongly coupled oxygen-17 ($I$=5/2) nuclei in the NHB hydroxyl groups dominate the EPR spectrum, the latter significantly deviates from the Gaussian. Only averaging over many cluster configurations can remove the structure (see Part 2 of this study). [7] For protons, even a single configuration of the anion cluster yields EPR spectra that are essentially Gaussian (provided that both the first and the second solvation shells are included).

For distant nuclei, in the point dipole approximation, $a \approx 0$ and

$$-T_\perp \approx \frac{g \beta_e g_{n,k} \beta_N}{r_{Xk}^3} ,  \qquad (A9)$$

where $r_{Xk}$ is the distance between the centroid of the electron wavefunction and the $k$-th nucleus; $g_{n,k}$ is the corresponding nuclear $g$-factor. For most of the nuclei of interest, including $^2$H (but not $^{17}$O), $g_n > 0$, $T_\perp < 0$ and thus $B_{zz} > 0$. It is seen from eq. (A9) that $|T_\perp|$ rapidly decreases with the distance, so the contribution $M_2^{(m)}$ of such distant (matrix) nuclei to the second moment is quite small. Angle averaging over the sample yields

$$M_2^{(m)} \approx \frac{4}{15} \sum_k \left(g_{n,k}\beta_N\right)^2 I_k(I_k+1) \left\langle r_{Xk}^{-6}\right\rangle . \tag{A10}$$

The last term in eq. (A10) can be estimated assuming uniform distribution of the nuclei (with the number density of $\rho_n$) beyond a sphere with a cutoff radius of $r_{cut}$: $\left\langle r^{-6}\right\rangle \approx 4\pi\rho_n / 3r_{cut}^3$. For $^{17}O$ nuclei, the point-dipole contribution to $T_\perp$ is negligibly small, due to the smallness of $g_n^O < 0$, relatively long $X$-$O$ distances, and large contribution from the unpaired electron density in the $O$ $2p$ orbitals. The latter can be estimated from $B_{zz}^O / g\beta_e \approx 4/5\,g_n^O\beta_N\langle r_O^{-3}\rangle$, where averaging is over the $np$ orbital of the oxygen (for the radius vector $\mathbf{r}_O$ centered on the oxygen-17 nucleus).

Eq. (A3) includes only first-order effects. EPR spectra of trapped electrons commonly exhibit additional satellite lines at $\Delta B \approx \pm\nu_H\left[1+9/32(T_\perp/\nu_H)^2\right]$ [29] due to ("forbidden") spin transitions that involve simultaneous flips of electron and nuclear spins. Such flip-flop satellites are spectrally resolved only in protiated samples. The ratio of the satellite to the central resonance line intensities is given by [29]

$$I_{sat}/I_{central} \approx 3/20 \ n(T_\perp/\nu_H)^2 , \tag{A11}$$

where $n$ is the number of magnetically equivalent protons in the first coordination shell of trapped electron. Eq. (A11) provides yet another constraint on $n$ and $T_\perp$.

## 2. Electron spin echo envelope modulation (ESEEM) spectra.

For a single nucleus coupled to the electron spin, the NMR transitions will occur with the frequencies $\nu_{\alpha,\beta}$ (which depend on the orientation of the electron spin up ($\alpha$) and down ($\beta$) the field, respectively): [33]

$$\nu_{\alpha,\beta} = \left[\left(\frac{A}{2}\mp\nu\right)^2 + \left(\frac{B}{2}\right)^2\right]^{1/2} , \tag{A12}$$

where (in our conventions)

$$A = a + T_\perp\left(1 - 3\cos^2\theta\right) , \tag{A13}$$

$$B = 3T_\perp\sin\theta\cos\theta , \tag{A14}$$

and $\theta$ is the angle between axis $z$ of the laboratory frame and the long axis of the hfcc tensor. The primary ESEEM (pESEEM) kinetics $V(\tau)$ are obtained by following the primary spin echo signal induced using a $\pi/2 - \tau - \pi - \tau$ sequence of two short microwave pulses as a function of the delay time $\tau$ between the pulses with the spin flip angles of $\pi/2$ and $\pi$. The pESEEM kinetics for a spin-1 nucleus ( a deuteron) are given by

$$V_k(\tau) = 1 - \frac{16}{3}\xi^2\left(1 - \xi^2\right), \qquad (A15)$$

where

$$\xi = \frac{\nu B}{\nu_\alpha \nu_\beta}\sin\left(\pi\nu_\alpha\tau\right)\sin\left(\pi\nu_\beta\tau\right).$$

For a system of $k$ nuclei,

$$V(\tau) = \prod_k V_k(\tau). \qquad (A16)$$

The pESEEM spectrum is a power Fourier Transform (FT) spectrum of the time domain $V(\tau)$ kinetics. The most prominent peaks in this spectrum correspond to frequencies given by $\nu_{\alpha,\beta}$ and $\nu_\alpha + \nu_\beta$. The first 100-200 ns of the pESEEM kinetics cannot be observed (due to the "dead time" of pulsed EPR spectrometer) and this introduces distortions in the resulting FT spectrum. Likewise, the occurrence of transverse relaxation (that to a first approximation can be taken into account by multiplying eq. (A7) by $\exp\left(-2\tau/T_2\right)$, where $T_2$ is the relaxation time) limits the observation window to ca. 4 μs, limiting the spectral resolution. Such effects have to be taken into account for comparison of the experimental and simulated data.

Nuclei with large hfcc constants and small spin (e.g., protons) introduce modulation that is too fast and has too small amplitude to provide reliable estimates for $^1$H hfcc's. For that reason, proton hfcc's are estimated from the data obtained for perdeuterated samples; the proton hyperfine constants are 6.5 times larger than the corresponding $^2$H constants. Importantly, not only the first solvation shell but also matrix nuclei contribute to $V(\tau)$. The contribution from these matrix protons dominates the stimulated ESEEM for $e_{hyd}^-$ [32,33] which, for that reason, is not particularly useful. Since the amplitude of the modulation increases as $(T_\perp/\nu_D)^2$, distant (matrix) nuclei do not yield large contribution to pESEEM. In the 9 GHz band, deuterons with $r_{XD} > 2.85$ Å can be considered as distant ones. [33] Such nuclei give sharp features in the pESEEM spectra at the modulation frequencies of $\nu_D$ and $2\nu_D$.

The kinetics given by eq. (A16) must be averaged over all orientations of the electron center with regard to the external field. If the coupling constants are relatively small, the two most prominent frequencies in such an angle averaged spectrum are

$$\nu_{\alpha(\beta)}^\perp \approx \left| \nu \mp \frac{a + T_\perp}{2} \right| \qquad (A17)$$

and

$$\nu_\alpha + \nu_\beta \approx 2\nu + \frac{T_\perp^2}{2\nu}. \qquad (A18)$$

For trapped electron in alkaline ice (the experimental FT pESEEM spectrum, before and after rejection filtering, is shown in Figure A2), the peak maxima for deuterons in the first solvation shell are at $\nu_{\alpha(\beta)}^{\perp} \approx 3.15$ MHz (peak (i)) and $\nu_{\alpha} + \nu_{\beta} \approx 4.9$ MHz (peak (ii); for $\nu_D \approx 2.2$ MHz), which gives $a \approx$ -0.4 MHz and $T_{\perp} \approx$ -1.5 MHz (assuming that $T_{\perp} < 0$); for protons, this corresponds to $a^H \approx -0.92$ G and $B_{zz}^H \approx +7$ G, respectively, in field units. The modulation frequencies from matrix nuclei that are weakly coupled to the electron spin are exhibited prominently at $\nu_D$ and $2\nu_D$ and were suppressed by rejection filtering. Despite that, these frequencies are still prominent. Figure A2 exhibits the simulated power FT pESEEM spectrum for these parameters obtained by Astashkin et al. [32] in the model with one or two magnetically equivalent deuterons coupled to the electron spin. [32,33] It is seen that the model accounts for the positions of the maxima of peaks (i) and (ii) but accounts poorly for their widths and does not simulate at all the weak, broad features with frequencies above 6 MHz (marked by asterisks in the spectrum). In Figure A3, the effect of a distribution in the spin parameters is illustrated, assuming a Gaussian distribution for $T_{\perp}$ (centered around –1.5 MHz) with a standard deviation of 0.4 MHz (a single-deuteron calculation). This results in broadening of the two main lines (corresponding to peaks (i) and (ii) in Figure A2) that is comparable to that experimentally observed, but it also fails to account for the high frequency part of the spectrum. As shown in Part 2 of this study, [7] simulating these features requires averaging over realistic deuteron configurations; it is impossible to find a single configuration (of one, two, or six magnetically equivalent deuterons) that is "representative" of all such configurations, even if one introduces *ad hoc* distributions for the spin parameters.

**Appendix B. Geometries of large water anion clusters.**

Optimized geometry (B(X)3LYP with 6-31+G* based basis set). The centroid is at the origin. The Cartesian coordinates are given in Angstroms.

*1. w20n2, n=20 cluster (Khan [20]); Figure 10S.*

```
O  1  -2.3319  1.0080 -2.8890
H  2  -1.3799  0.7920 -2.8040
H  3  -2.4470  1.7789 -2.3020
O  4  -3.9169  -.7580  0.7940
O  5   0.3010  0.3350 -2.2560
O  6   3.4249 -1.9929 -1.3289
O  7   1.2909  3.3309  1.8140
H  8  -3.3590 -1.2869  1.3910
H  9  -3.6770 -1.0280  -.1110
H 10   0.6160  -.5320 -2.6390
H 11   0.1900  0.1580 -1.2929
O 12  -3.3399 -1.3209 -1.9960
H 13   3.5779 -1.0269 -1.1390
H 14   4.2649 -2.3149 -1.6710
O 15   3.9590  0.6170  -.8550
H 16   1.4230  4.1440  2.3130
H 17   0.9810  3.6070  0.9090
O 18   1.1339 -2.0529 -3.1310
H 19  -3.0190  -.4390 -2.3290
H 20  -4.1390 -1.5129 -2.4969
H 21   3.4110  1.2460 -1.3559
H 22   3.8609  0.8930  0.0850
O 23   2.4289 -3.2650  1.0709
O 24   -.7610  1.4199  2.2789
O 25   1.9150  -.9170  2.4710
O 26   -.5040  -.7870  4.0300
O 27  -2.2210 -2.3359  2.6400
H 28   1.8810 -2.2240 -2.5310
H 29   0.3830 -2.6179 -2.8410
O 30  -1.1770 -3.3170 -2.1579
H 31   1.4780 -3.3700  0.8850
H 32   2.8040 -2.9050  0.2460
H 33   -.5240  0.8420  1.5150
H 34   -.0830  2.1269  2.2549
O 35   3.5060  1.3790  1.7980
H 36   2.2469 -1.7810  2.1330
H 37   1.3320  -.6200  1.7400
H 38   0.4   -.9900  3.7199
H 39   -.6650  0.0940  3.6370
H 40  -2.6829 -2.9260  3.2450
H 41  -1.6099 -1.7840  3.2029
H 42  -1.9070 -2.6749 -2.2099
H 43   -.9710 -3.3490 -1.2030
H 44   2.8109  2.0590  1.8120
H 45   3.0830  0.5860  2.1779
O 46   -.3970 -2.8910  0.5260
O 47   0.4010  3.9699  -.6400
O 48   -.2490 -1.9280  0.3810
H 49  -1.0489 -2.9060  1.2549
O 50  -3.3020  1.9580  1.1690
H 51   -.5160  3.6669  -.7870
H 52   0.9600  3.4800 -1.2740
```

```
H 53  -2.4880 1.8599 1.7029
H 54  -3.6019 1.0329 1.0249
O 55   2.0569 2.3300 -2.3029
H 56   2.2589 2.5920 -3.2080
H 57   1.4130 1.5720 -2.3700
O 58  -2.3810 3.1690 -1.0029
H 59  -2.7530 2.7119 -.1910
H 60  -2.9449 3.9360 -1.1490
```

### *2. t24n1, n=24 cluster (Khan [20]); Figure 11S.*

```
O  1   -1.1458 2.9398 -2.8589
H  2   -1.4659 3.3900 -2.0590
H  3   -1.4929 2.0299 -2.7829
O  4   -.2770 -2.9030 2.9750
O  5   1.9170 -3.9729 -1.0668
H  6   -.2500 -3.0120 2.0099
H  7   0.6119 -2.5510 3.1890
O  8   1.6350 2.6169 -2.8799
O  9   -2.3109 1.2370 3.1290
O  10  4.3098 -.2620 -.7159
H  11  1.0908 -3.6150 -.6400
H  12  1.8249 -4.9329 -1.0500
O  13  -2.7090 -3.5400 -1.0968
H  14  1.9489 3.0539 -2.0720
H  15  0.6660 2.7800 -2.9190
H  16  -2.8849 1.6009 2.4359
H  17  -2.4919 0.2690 3.1699
O  18  -1.9969 0.3019 -2.1930
H  19  4.1798 -1.1120 -.2540
H  20  3.5499 -.1900 -1.3369
O  21  2.4260 -1.9750 3.1669
H  22  -3.3319 -3.0259 -.5270
H  23  -2.9889 -4.4610 -1.0448
O  24  3.7900 -2.7660 0.5829
H  25  -1.3389 0.1529 -1.4830
H  26  -1.9030 -.5 -2.7690
H  27  2.8449 -2.2259 2.3300
H  28  2.5160 -.9989 3.2080
O  29  0.1320 1.5300 2.0379
O  30  2.7829 3.7930 -.4460
O  31  -4.3258 0.3469 -.6650
H  32  3.1789 -3.2480 -.0200
H  33  4.5240 -3.3669 0.7500
O  34  -2.7160 -1.4959 3.1230
O  35  1.1518 -2.4279 -3.5300
O  36  -3.7730 2.6360 0.9669
H  37  -.7319 1.4110 2.5169
H  38  0.0950 0.9010 1.2940
H  39  3.3319 3.1189 0.0219
H  40  3.3730 4.5270 -.6420
H  41  -4.1958 1.1378 -.1110
H  42  -3.5709 0.3509 -1.2948
H  43  -3.1269 -1.7520 2.2829
H  44  -1.8540 -1.9670 3.1589
H  45  1.4899 -3.0230 -2.8420
H  46  1.4950 -1.5510 -3.2669
O  47  -.2379 -2.8689 0.0729
H  48  -3.1339 3.2309 0.5070
H  49  -4.4729 3.2069 1.2980
O  50  -1.6590 -2.0929 -3.4870
O  51  2.6410 0.8369 3.1349
H  52  1.9910 -.0140 -2.2400
O  53  -4.2279 -1.9809 0.5739
H  54  -1.0948 -3.1139 -.3380
H  55  -.1640 -1.8900 -.0420
```

```
H 56  -1.9950 -2.6949 -2.8060
H 57  -.6929 -2.2589 -3.5489
H 58  3.1989 1.1858 2.4209
H 59  1.7359 1.1140 2.8900
O 60  4.2488 1.8460 0.8469
H 61  1.3128 -.0929 -1.5380
H 62  1.8820 0.9150 -2.5750
H 63  -4.2750 -1.0768 0.1409
H 64  -5.1360 -2.2130 0.7950
O 65  -1.9419 4.2488 -.3459
O 66  0.3069 4.0650 0.9849
H 67  4.2880 1.0280 0.2620
H 68  5.1580 2.0270 1.1080
H 69  -1.0628 4.2110 0.1439
H 70  -2.1440 5.1840 -.4550
H 71  1.1458 4.0579 0.4919
H 72  0.2650 3.1840 1.4290
```

### 3. $4^6 6^8 B$, n=24 cluster (Herbert & Head-Gordon [21]); Figure 12S.

```
O  1   -4.735642     .228668    -1.320304
O  2   -2.254896     .099324    -2.512210
O  3    -.626335   -1.903906    -3.715820
O  4    -.418905   -3.984856    -2.031394
O  5   -2.197186   -3.759506     .220653
O  6   -4.268045   -1.987205     .507184
O  7   -4.078217    2.358259     .563832
O  8   -3.940922     .150583    2.441015
O  9     .095029   -4.065446    1.966168
O 10    1.892614   -3.996367     -.293637
O 11    1.210602    -.003995    -2.701064
O 12    -.506782    2.028624    -3.666741
O 13    4.735554     -.228420    1.320257
O 14    4.267866    1.987113     -.507472
O 15    2.197078    3.759673     -.221280
O 16     .419008    3.984647    2.030883
O 17     .626530    1.904404    3.715753
O 18    2.255093     -.099232    2.512848
O 19    4.078285   -2.358316     -.563724
O 20    3.941069     -.150769   -2.441083
O 21    -.095171    4.066058    -1.966661
O 22   -1.892595    3.996286     .293356
O 23   -1.210433     .004368    2.701632
O 24     .506788   -2.028587    3.667262
H 25   -3.445104   -2.558282     .369198
H 26   -1.681630   -3.798783     -.627912
H 27    -.509106   -3.208817   -2.683138
H 28   -1.396444   -1.337732   -3.447365
H 29   -3.219875     .161671   -2.239150
H 30   -4.647790     -.572817     -.743476
H 31   -4.022448     .957015    1.869395
H 32    1.383868   -4.007719     .559707
H 33     .234755    1.393141   -3.472863
H 34    4.647393     .572994     .743375
H 35    3.444984    2.558320     -.369671
H 36    1.681536    3.798835     .627308
H 37     .509247    3.208739    2.682802
H 38    1.396640    1.337975    3.447827
H 39    3.220044     -.161555    2.239705
H 40    4.022627     -.957230   -1.869510
H 41   -1.383935    4.007899     -.560042
H 42    -.234708   -1.393063    3.473373
H 43   -4.579915    1.003769     -.722621
H 44     .155696   -1.326859   -3.500529
H 45   -1.501293   -3.796487     .927248
H 46   -3.224533    2.883933     .427428
H 47     .264480   -3.308121    2.623567
H 48    2.215917     -.050077   -2.740411
H 49    4.107238     .619077   -1.838558
H 50   -1.188729    3.914779     .988938
H 51    1.310676   -1.515746    3.389856
H 52   -4.107090     -.619293    1.838532
H 53    1.188763   -3.914952     -.989230
H 54   -1.310613    1.515734   -3.389274
H 55    3.224509   -2.883905     -.427577
```

```
H 56     -.264742      3.308592     -2.623891
H 57    -2.215770       .050110      2.740739
H 58     4.579769     -1.003576       .722664
H 59     1.501250      3.797010      -.927925
H 60     -.155543      1.327160      3.501133
H 61    -1.768839       .056959     -1.648792
H 62     -.515491     -4.801086     -2.564646
H 63    -4.992008     -2.623531       .681768
H 64    -4.763290      3.031175       .757444
H 65      .091207     -4.890340      2.495179
H 66     1.001099      -.012518     -1.730289
H 67     4.763365     -3.031308      -.757064
H 68     -.091892      4.890899     -2.495744
H 69    -1.741         .012579      1.730892
H 70     4.991998      2.623301      -.681847
H 71      .515526      4.800984      2.563976
H 72     1.769         -.057116      1.649434
```

### 4. $5^{12}6^2B$, n=24 cluster (Herbert & Head-Gordon [21]); Figure 13S.

```
O  1   4.129436    2.459760   -0.764238
O  2   2.358367    4.192957    0.710200
O  3   0.047190    4.540578   -0.764889
O  4  -2.459276    4.128955    0.764640
O  5  -4.193045    2.358375   -0.709803
O  6  -4.541329    0.047361    0.765546
O  7  -4.129597   -2.459445   -0.763518
O  8  -2.357975   -4.192774    0.710298
O  9  -0.047087   -4.540557   -0.765226
O 10   2.459892   -4.129169    0.763531
O 11   4.193094   -2.357977   -0.710831
O 12   4.541093   -0.047110    0.764603
O 13   2.703580    1.446522   -3.237602
O 14   0.066556    2.103381   -2.293337
O 15  -2.665963    1.473943   -3.181353
O 16  -2.704447   -1.446407   -3.237252
O 17  -0.067246   -2.103331   -2.293551
O 18   2.665170   -1.473752   -3.181775
O 19  -1.473378   -2.665679    3.181599
O 20  -2.103746    0.066548    2.293332
O 21  -1.446033    2.703182    3.237854
O 22   1.474253    2.665134    3.181427
O 23   2.103723   -0.066997    2.292708
O 24   1.446987   -2.703830    3.237327
H 25   2.722599    0.489622   -3.280899
H 26   0.892079    1.919939   -2.749015
H 27  -1.745157    1.698793   -3.056187
H 28  -2.723481   -0.489510   -3.280593
H 29  -0.892865   -1.919962   -2.749086
H 30   1.744404   -1.698698   -3.056488
H 31  -1.698459   -1.744935    3.056390
H 32  -1.919994    0.891952    2.749104
H 33  -0.489129    2.722234    3.281049
H 34   1.699205    1.744401    3.055886
H 35   1.920336   -0.892504    2.748442
H 36   0.490092   -2.722970    3.280684
H 37   3.179830    1.687393   -2.448738
H 38  -3.107655    1.772535   -2.391883
H 39  -0.050386   -1.485903   -1.559526
H 40  -3.801898    0.043478    1.376686
H 41   2.615763    5.059606    0.995590
H 42   3.801685   -0.043317    1.375771
H 43   4.306197   -1.534935   -0.216210
H 44   3.010079   -3.518918    0.268514
H 45   0.756422   -4.455836   -0.257596
H 46  -1.534964   -4.305741    0.215594
H 47  -3.519256   -3.009531   -0.268499
H 48  -4.456770   -0.756097    0.257808
H 49  -4.306263    1.535329   -0.215221
H 50  -3.009604    3.518971    0.269452
H 51  -0.756192    4.455897   -0.257051
H 52   1.535275    4.305900    0.215627
H 53   4.844437    3.024372   -1.025867
H 54  -1.772108   -3.107505    2.392256
H 55  -1.686991    3.179453    2.449028
H 56   1.486161   -0.050176    1.558796
```

```
H 57   0.043309    3.801237   -1.376136
H 58  -4.844580   -3.024193   -1.024905
H 59   3.024427   -4.844448    1.024570
H 60   5.059662   -2.615403   -0.996440
H 61   0.049704    1.485885   -1.559371
H 62  -3.180510   -1.687237   -2.448261
H 63   3.107049   -1.772431   -2.392443
H 64  -1.486417    0.049674    1.559227
H 65   1.772695    3.107162    2.392090
H 66   1.687870   -3.180129    2.448497
H 67   4.456547    0.756419    0.256978
H 68   3.519320    3.009965   -0.269074
H 69  -3.023705    4.844156    1.026120
H 70  -5.059618    2.616122   -0.995108
H 71  -2.615446   -5.059442    0.995562
H 72  -0.043404   -3.801219   -1.376477
```

**Table 1S.**

Comparison of different computational methods: Relative energies and isotropic hfcc's for $^1H$ and $^{17}O$ nuclei in water tetramer anions (optimized geometry).

| water tetramer anion | $a$, G | BLYP 6-31+G** | B3LYP 6-31+G** | B3LYP 6-311++G** | B3LYP aug-cc-VDZ |
|---|---|---|---|---|---|
| $C_{4h}$ OH-type | $H_a$ | 4.4 | 2.9 | 0.71 | 0.61 |
| Fig. 1S(d) | O | -77.7 | -75.4 | -24.3 | -17.6 |
| | $H_b$ | 8.55 | 5.5 | 0.68 | 0.78 |
| energy, meV | | *710* | *796* | *0* | *0* |
| $C_{4h}$ b-type | $H_a$ | 4.3 | 0.2 | -0.65 | -0.37 |
| Fig. 1S(c) | O | -52.7 | -51.8 | -24.9 | -21.7 |
| | $H_b$ | 1.35 | 0.45 | 0.22 | 0.52 |
| energy, meV | | *260* | *290* | *257* | *290* |
| $D_{2d}$ b-type | $H_a$ | 3.32 | -0.67 | -0.7 | -0.45 |
| Fig. 1S(a) | O | -45.3 | -43.9 | -22.4 | -19.5 |
| | $H_b$ | 0.28 | 0.34 | 0.15 | 0.14 |
| energy, meV | | *125* | *130* | *262* | *287* |
| $D_{2d}$ d-type | $H_a$ | 0.06 | -1.2 | -0.83 | -0.46 |
| Fig. 1S(b) | O | -41.2 | -39 | -19 | -17.6 |
| | $H_b$ | 0.8 | -1.56 | -1.08 | -0.68 |
| energy, meV | | *0* | *0* | *104* | *152* |

a)  the energy of the most stable isomer is taken for zero.

**Table 2S.**

Comparison of isotropic hfcc's (*a*) for $^1$H and $^{17}$O nuclei and the shortest XH distances for $D_{2d}$ bond type water tetramer anions (optimized geometries; Figure 1S(a)) for different methods and basis sets.

| basis set | a, G | HF | BLYP | B3LYP | LSDA | MP2 |
|---|---|---|---|---|---|---|
| 6-31+G** | H$_a$ | 9.4 | 3.32 | -0.67 | 4.5 | 12.3 |
| | O | -35.1 | -45.3 | -43.9 | -46 | -38.9 |
| | H$_b$ | 1.6 | 0.28 | 0.34 | -0.58 | 2.13 |
| $r_{XH}$, Å | | *1.74* | *1.55* | *1.57* | *1.36* | *1.57* |
| 6-311++G** | H$_a$ | -1.7 | 1.1 | -0.7 | 1.65 | -3.9 |
| | O | -7.02 | -23.1 | -22.4 | -31.1 | -13 |
| | H$_b$ | -0.4 | 0.39 | 0.15 | 0.12 | -0.72 |
| $r_{XH}$, Å | | *2.94* | *1.89* | *1.96* | *1.47* | *2.17* |
| aug-cc-pVDZ | H$_a$ | -1.2 | 1.28 | -0.45 | 2.17 | -4.05 |
| | O | -4.9 | -21.5 | -19.5 | -29.1 | -13.4 |
| | H$_b$ | -0.24 | 0.73 | 0.14 | 0.11 | -0.61 |
| $r_{XH}$, Å | | *3.2* | *1.89* | *1.96* | *1.45* | *1.98* |
| EPR-III | H$_a$ | -1.68 | 1.46 | | | |
| | O | -6.16 | -25.1 | | | |
| | H$_b$ | -0.44 | 0.93 | | - | - |
| $r_{XH}$, Å | | *3.12* | *1.98* | | | |

**Table 3S.**

Hyperfine coupling constants, second moments of EPR spectra, and spin and charge densities for selected $n=20$ and $n=24$ water anion clusters (B3LYP/6-311++G**)

| $n$ | 20 | 24 | 24 | 24 |
|---|---|---|---|---|
| cluster | *w20n2* [a] | *t24n1* [a] | $4^6 6^8 B$ [b] | $5^{12} 6^2 B$ [b] |
| Figure | 9S [d] | 10S | 10S | 12S |
| $a^O$ (1), G | -25 | -22.8 | -17.9 | -16.3 |
| $a^H$ (1), G | 0.19 | 0.15 | 0.32 | -0.65 |
| $a^O$ (2), G | -3.1 | -2.5 | -1.5 | -1.53 |
| $a^H$ (2), G | -0.003 | 0.025 | 0.12 | 0.03 |
| $M_2^O$, G$^2$ | 8340 | 8180 | 5200 | 4190 |
| $M_2^H$, G$^2$ | 27.2 | 20.8 | 8.5 | 9.7 |
| $iso$ [c] | 0.9 | 2.3 | 1.1 | *0.98* |
| $B_{zz}^H$ (1), G | 6.8 | 5.6 | 3.5 | 3.8 |
| $B_{zz}^O$ (1), G | -3.1 | -2.5 | -1.5 | -1.5 |
| $\rho_s^O$ (1) | -0.067 | -0.045 | -0.048 | -0.017 |
| $\rho_s^H$ (1) | -0.11 | -0.05 | -0.044 | -0.088 |
| $\rho_s^O$ (2) | 0.006 | 0.001 | -0.005 | 0.004 |
| $\rho_s^H$ (2) | 0.019 | 0.014 | 0.019 | 0.014 |
| $\rho_c^O$ (1) | -0.57 | -0.56 | -0.63 | -0.58 |
| $\rho_c^H$ (1) | 0.22 | 0.17 | 0.23 | 0.21 |
| $\rho_c^O$ (2) | -0.64 | -0.64 | -0.63 | -0.64 |
| $\rho_c^H$ (2) | 0.32 | 0.32 | 0.31 | 0.32 |
| $\phi_{2p}^O$ | 0.17 | 0.14 | 0.10 | 0.10 |
| type | tetr. | tetr. | square | tetr. |
| $\langle r_{XH} \rangle_{av}$, Å | 1.78 | 1.87 | 2.21 | 2.16 |

$a^{O,H}$ and $B_{zz}^{O,H}$ are isotropic and the $zz$ component of anisotropic hfcc's for $^1$H and $^{17}$O nuclei, respectively; (1) refers to NHB hydroxyl groups and (2) refers to all other nuclei; $\rho_{s,c}^{O,H}$ stands for atomic spin and charge density determined using Mulliken population analsysis; $\phi_{2p}^O$ is the total population of $O\ 2p$ orbitals.

(a) Khan; [20]
(b) Herbert and Head-Gordon. [21] See Appendix B for atomic coordinates.
(c) Contribution from isotropic hfcc to $M_2^H$.
(d) A simulated proton EPR spectrum for this water anion is shown in Figure A1(a) in the Supplement.

**Table 4S.**

Hypefine coupling constants and second moments of EPR spectra for selected $n=20$ and $n=24$ water anion clusters: comparison of computation methods.

| method | B3LYP | HF | B3LYP | HF |
|---|---|---|---|---|
| basis set | 6-31+G** | | 6-311++G** | |
| cluster | *w20n2* ($n=20$) | | | |
| $-a^O$ (1), G | 24.5 | 18.8 | 25 | 20.3 |
| $a^H$ (1), G | 0.46 | -8.7 | 0.19 | -8.6 |
| $-a^O$ (2), G | 3.3 | 2.1 | 4.0 | 1.23 |
| $a^H$ (2), G | 0.02 | -0.2 | -.003 | -0.19 |
| $M_2^O$, G² | 7690 | 4400 | 8390 | 5150 |
| $M_2^H$, G² | 32 | 124 | 27.2 | 122 |
| cluster | *t24n1* ($n=24$) | | | |
| $-a^O$ (1), G | 23.6 | 18.2 | 22.8 | 19.8 |
| $a^H$ (1), G | 0.44 | -8.0 | 0.15 | -8.0 |
| $-a^O$ (2), G | 2.9 | 0.74 | 2.5 | 0.91 |
| $a^H$ (2), G | 0.02 | -0.15 | 0.025 | -0.15 |
| $M_2^O$, G² | 7913 | 4311 | 8180 | 5090 |
| $M_2^H$, G² | 27.2 | 104 | 20.8 | 103 |
| cluster | *$4^6 6^8 B$* ($n=24$) | | | |
| $-a^O$ (1), G | 18.9 | 12.2 | 17.9 | 13.4 |
| $a^H$ (1), G | 0.72 | -1.06 | 0.32 | -5.2 |
| $-a^O$ (2), G | 3.85 | 5.8 | 1.5 | 1.3 |
| $a^H$ (2), G | 0.12 | -0.17 | 0.12 | -0.07 |
| $M_2^O$, G² | 5138 | 1920 | 5200 | 2310 |
| $M_2^H$, G² | 11.8 | 51.8 | 8.8 | 42.8 |
| cluster | *$5^{12} 6^2 B$* ($n=24$) | | | |
| $-a^O$ (1), G | 17.2 | 11.7 | 16.3 | 12.6 |
| $a^H$ (1), G | -0.36 | -6.0 | -0.65 | -5.2 |
| $-a^O$ (2), G | 3.2 | 0.81 | 1.53 | 0.96 |
| $a^H$ (2), G | 0.02 | -0.19 | 0.03 | -0.1 |
| $M_2^O$, G² | 4120 | 1650 | 4190 | 1940 |
| $M_2^H$, G² | 12.7 | 55.2 | 9.7 | 25.8 |

see the caption to Table 3S.

**Figure Cations (Supplement)**

**Figure A1**

Simulated EPR spectra for **w20n2** ($n$=20) water cluster anion (section 4.2): (a) for $^1H_2^{16}O$ water; (b) for $^1H_2^{16}O/^1H_2^{17}O$ water (37 at % $^{17}O$). The dots are the histogram of resonance field offsets obtained using eq. (A3), the red solid line are convoluted spectra, the green line is the first derivative (to the right) and the black line is the Gaussian fit.

**Figure A2**

Experimental and simulated modulo FT pESEEM spectra (X-band) of trapped electron in low-temperature alkaline glass (10 M NaOD:$D_2O$). Replotted data from ref. 33 (Figure 3 on p. 247). The modulation pattern is from deuterons. The vertical lines indicate $\nu_I$ (ca. 2.2 MHz) and the second and third harmonics. The pink line is the unfiltered spectrum, the red line is the spectrum after rejection filtering at 2.2, 4.4, and 7.6 MHz. The blue and green lines are simulations of Astashkin et al. [32] for hfcc parameters given in the text (for one and two magnetically equivalent deuterons, respectively). Observe that the broad high frequency features indicated by asterisks are not reproduced in these simulation.

**Figure A3**

The effect of distribution of the anisotropic hfcc coupling on FT pESEEM spectra (simulation). The thin lines give the spectrum as calculated, thick lines show the spectrum after taking into account the effect of finite time window and dead time (200 ns). See the legend in the figure; the simulation parameters are given in Appendix A.

**Figure 1S**

Isodensity maps for singly occupied molecular orbit (SOMO) of (a) $b$-type and (b) $d$-type $D_{2d}$ symmetrical tetrahedral water anions, and planar C4h symmetrical (c) $b$-type and (d) H-bonded square water anions. In (b-d), $\pm0.03$ $a_0^{-3}$ isodensity surfaces are shown; in (a) several isodensity surfaces are plotted together. Red is for positive, purple is for negative. (B3LYP/6-311++G** model, optimized geometries).

**Figure 2S**

(a) Total population $\phi_{2p}^O$ of oxygen $2p$ orbitals (filled circles, to the left) as a function of $r_{XH}$, the X-$H_a$ distance (B3LYP/6-311++G** calculation for the $D_{2d}$ symmetrical $b$-type water anion shown in Figure 1S(a)). (b) Mulliken population analysis: atomic charge (filled symbols, top) and spin (empty symbols, bottom) densities for $H_a$ (circles), $H_b$ (triangles) and O (squares).

**Figure 3S**

(a) Isotropic hfcc's for $^{17}O$ (filled squares; to the top) and $^1H$ (bottom) nuclei (the $H_a$ [filled circles] and $H_b$ [empty triangles] nuclei are shown together) vs. X-O and X-$H_{a,b}$ distances, respectively. Solid lines are exponential fits (B3LYP/6-311++G** calculation for *b*-type tetrahedral water anion); (b) The same as (a), for the *zz* principal component of anisotropic hfcc tensor. The solid line is the estimate obtained in the point-dipole approximation, eq. (A9) in Appendix A.

**Figure 4S**

As Figure 3S: the comparison of EPR parameters calculated using HF (empty symbols) and B3LYP (filled symbols) methods (see the legend in the plot) using 6-311++G** basis set. In (a), isotropic hfcc constants obtained using B3LYP method are plotted to the left and the constants obtained using B3LYP method are plotted to the right. The solid lines are exponential fits.

**Figure 5S**

The plots of contributions to second moment of EPR spectrum from (top) oxygen-17 and (bottom) protons (*b*-type tetrahedral water anion). See eq. (A7) in Appendix A. Empty symbols are for HF and filled symbols are B3LYP (6-311++G** basis set). The solid lines are guides for the eye.

**Figure 6S**

As Figure 4S, for d-type tetrahedral anions shown at the top of the figure. Observe the strong deviations from the point dipole approximation in (d).

**Figure 7S**

The structure of cube shaped $C_{4h}$ symmetrical *d*-type water octamer anion (B3LYP/6-311++G** model; optimized geometry).

**Figure 8S**

As Figure 4S, for the two cube shaped octamer water anions (*b*-type and *d*-type) shown in Figures 1(b) and 7S, respectively. See the legends in the plot.

**Figure 9S**

As Figure 1, for the ***w20n2*** anion (EPR parameters, for optimized geometry given in Appendix B, are given in Tables 3S and 4S). The density levels are given in units of $a_0^{-3}$ (red is for positive, purple is for negative).

**Figure 10S**

As Figure 9S, for the ***t24n1*** anion.

**Figure 11S**

As Figure 9S, for the ***$4^6 6^8 B$*** anion.

**Figure 12S**

As Figure 9S, for the $\boldsymbol{5^{12}6^2B}$ anion.

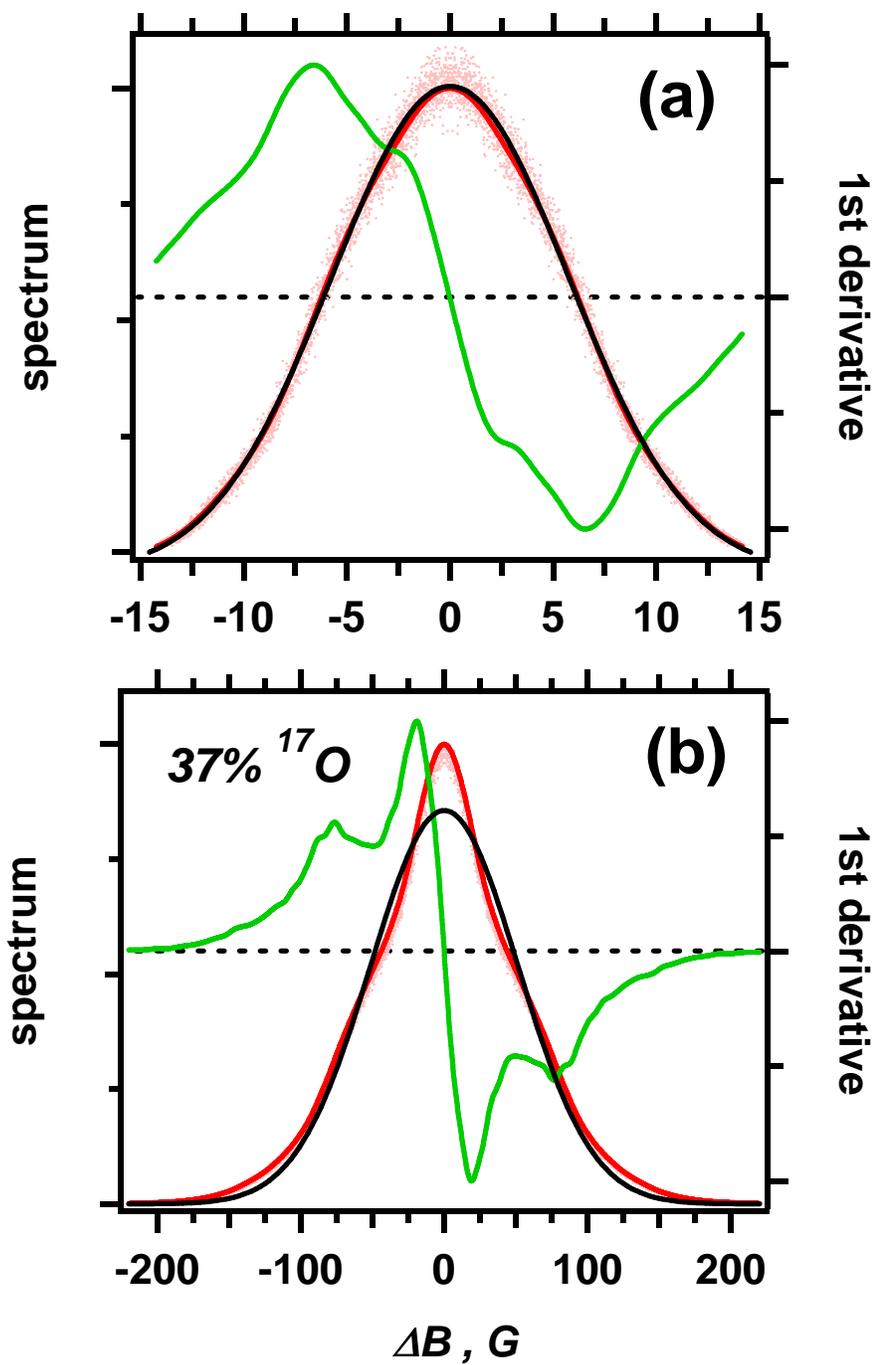

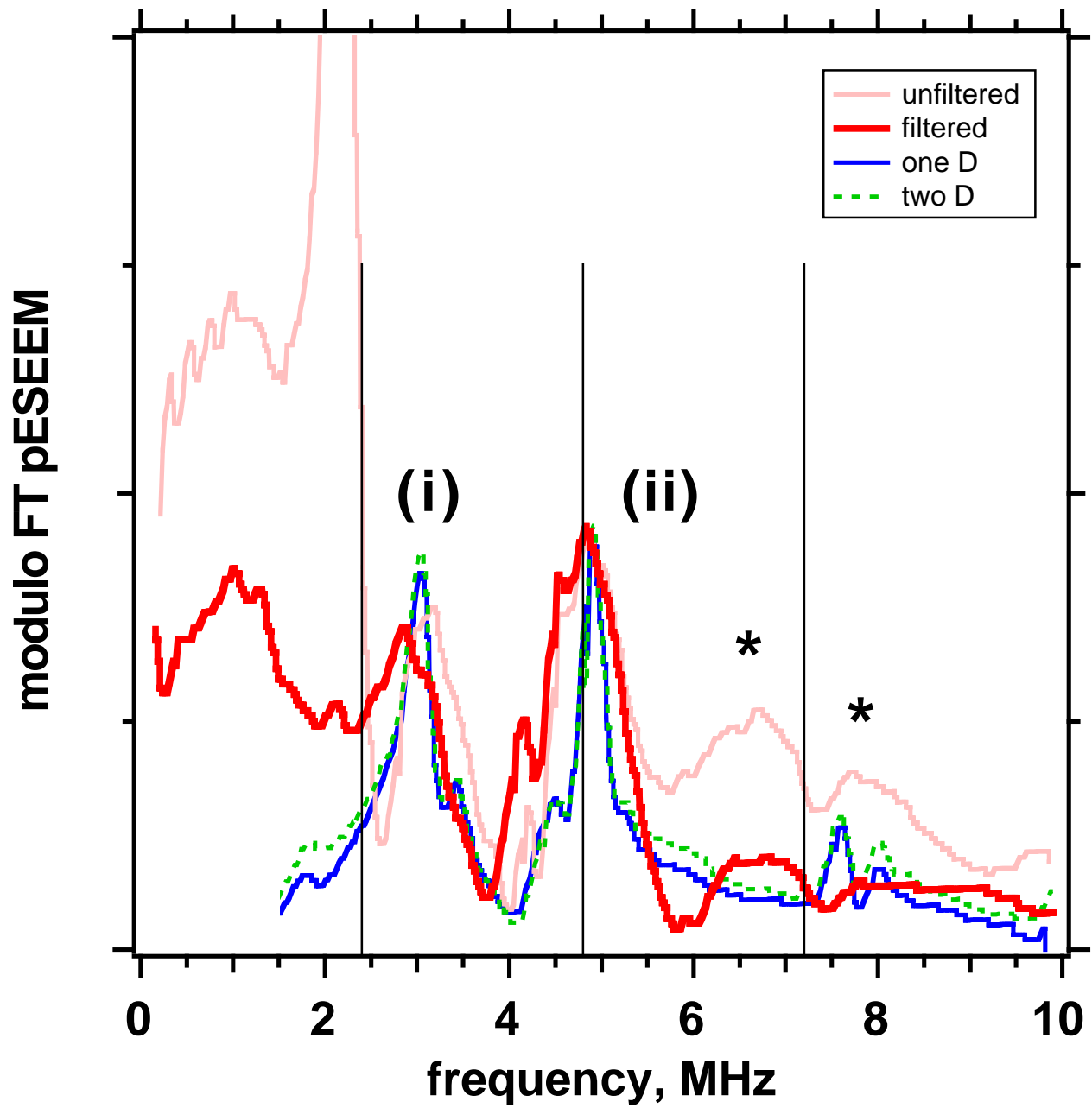

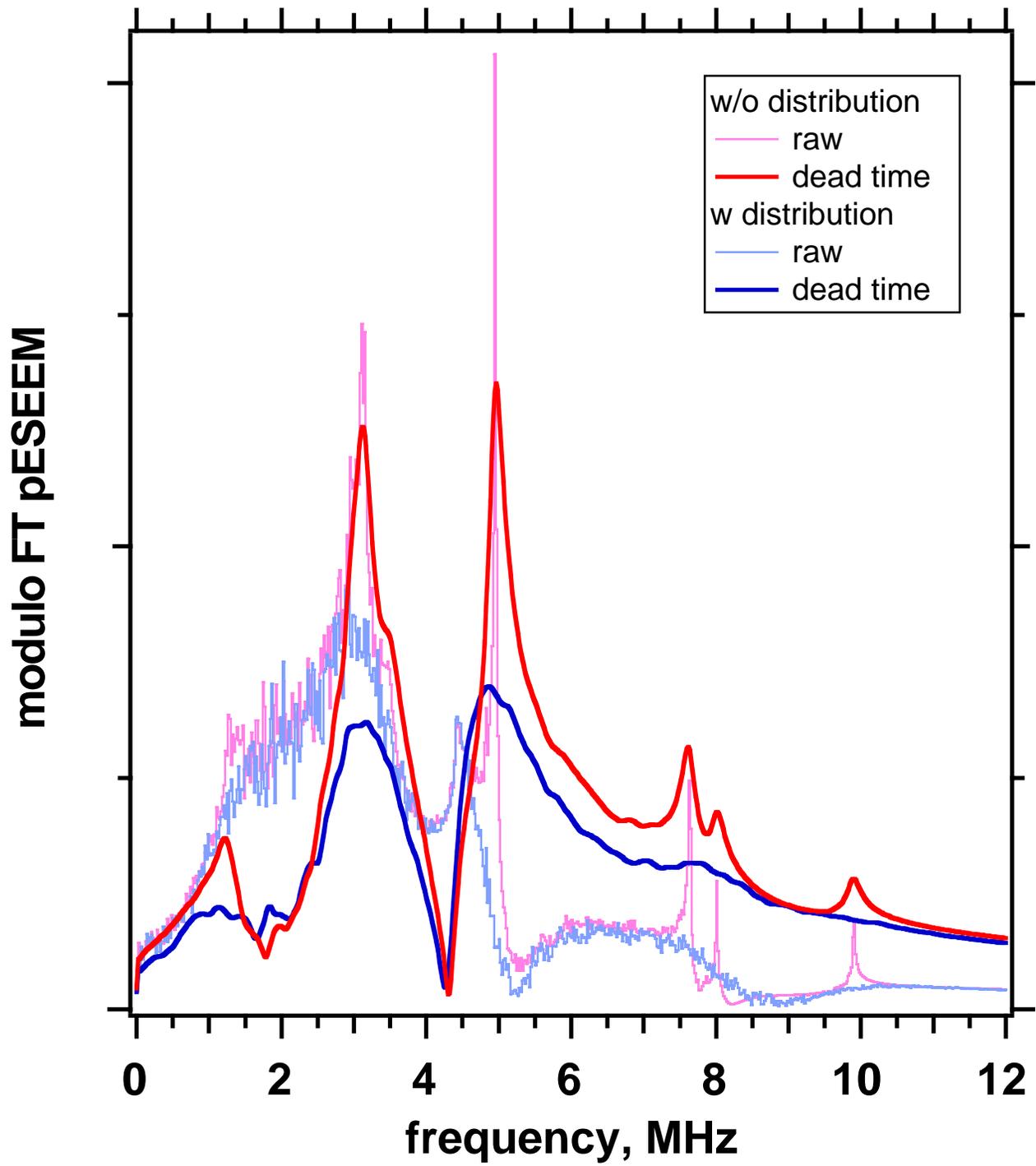

**n=4 water anion clusters**

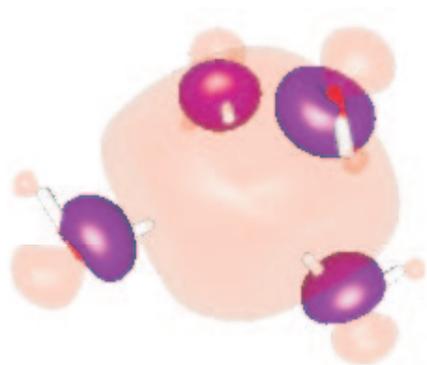

±0.03

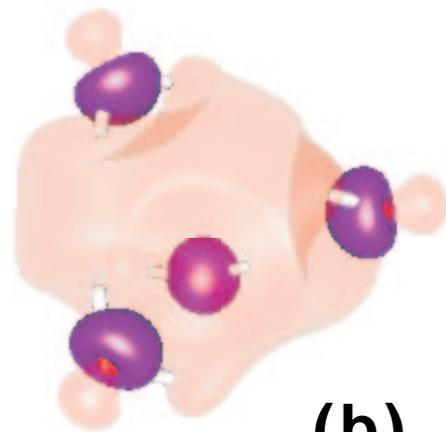

**(b)**

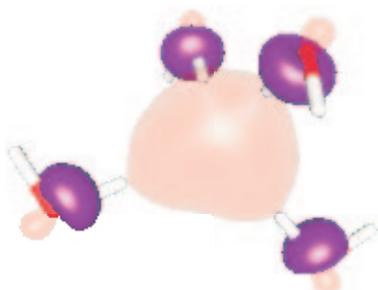

±0.06

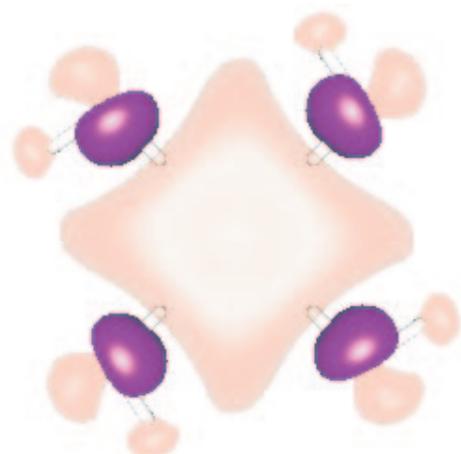

**(c)**

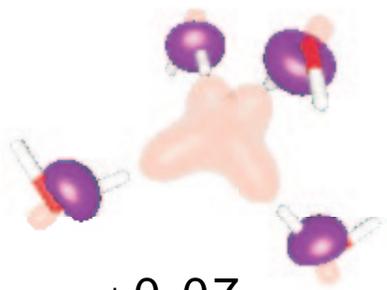

±0.07

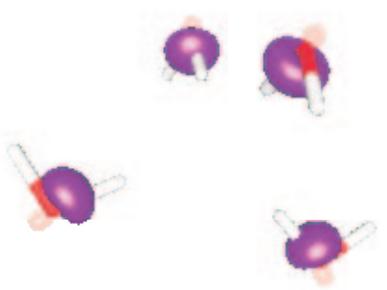

±0.08

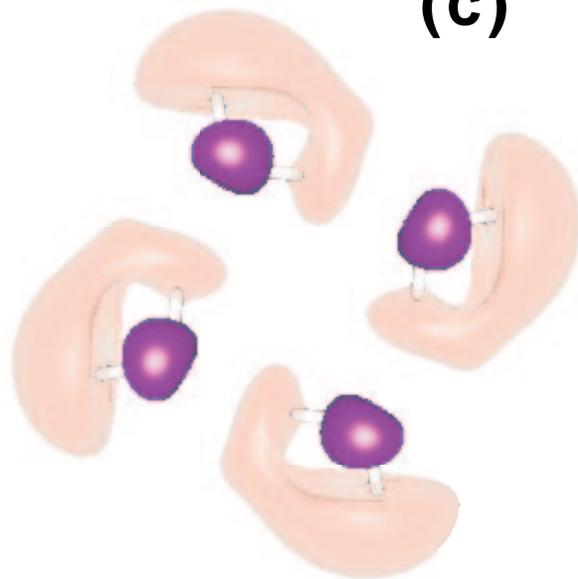

**(a)**

**(d)**

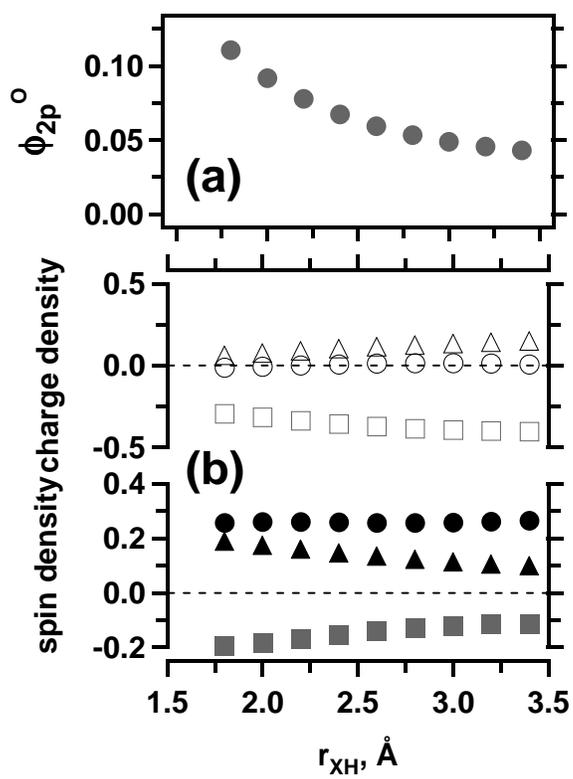

Figure 3S; Shkrob

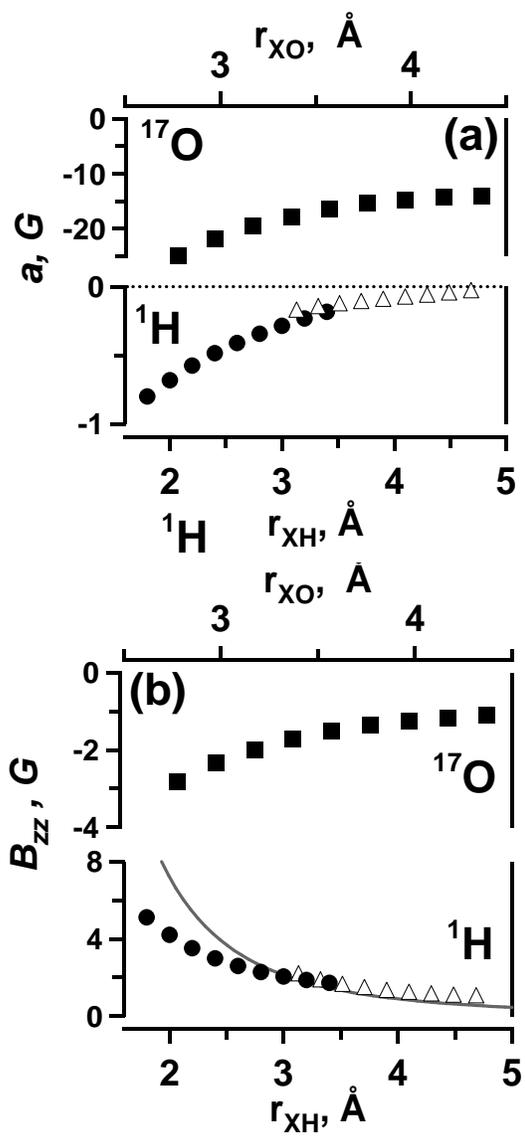

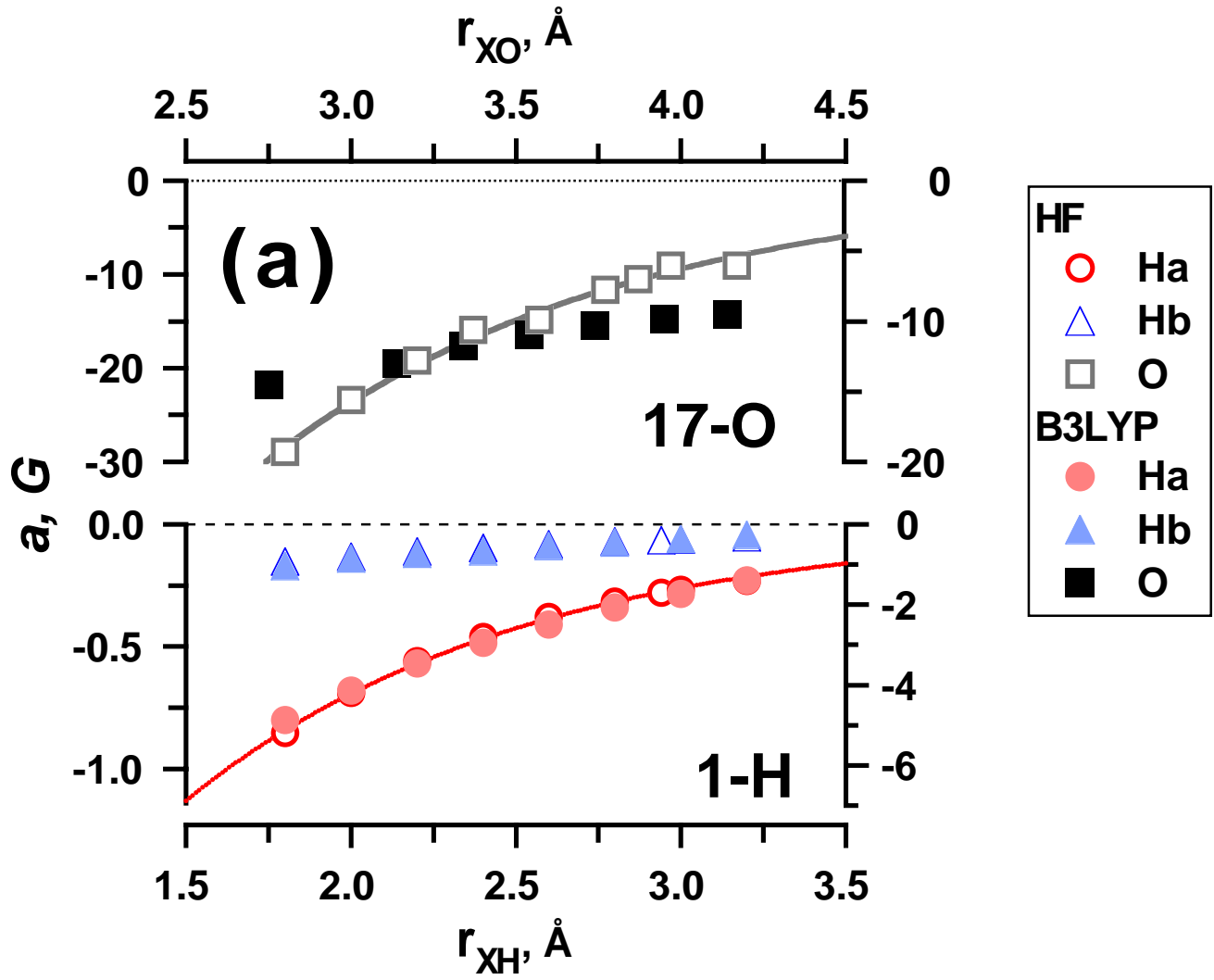

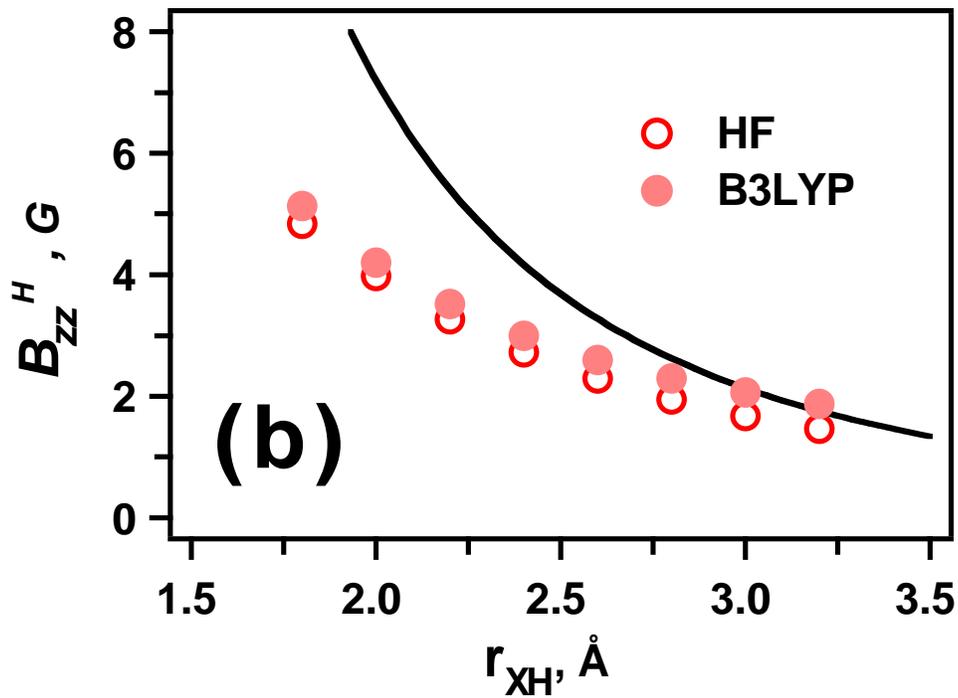

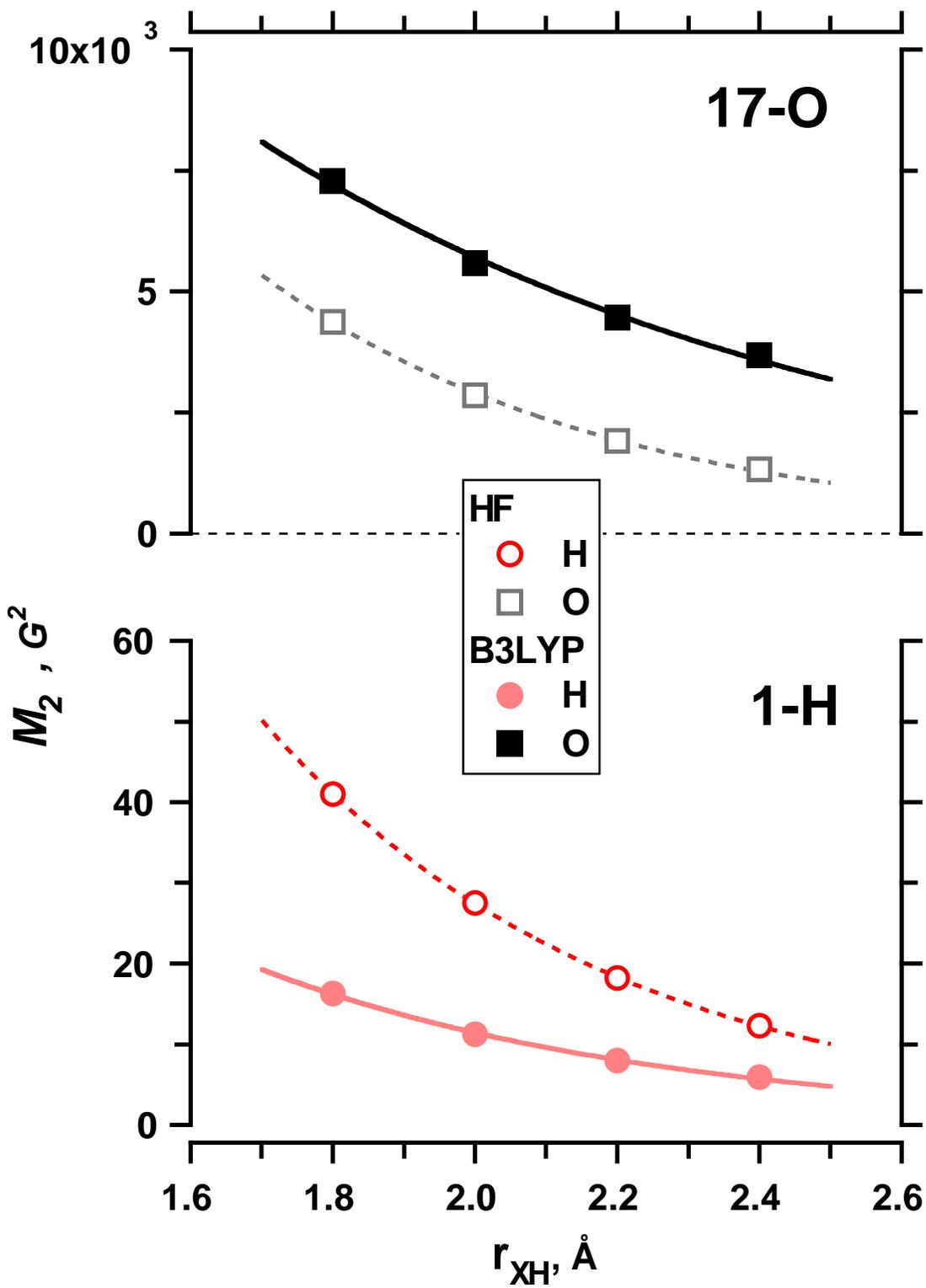

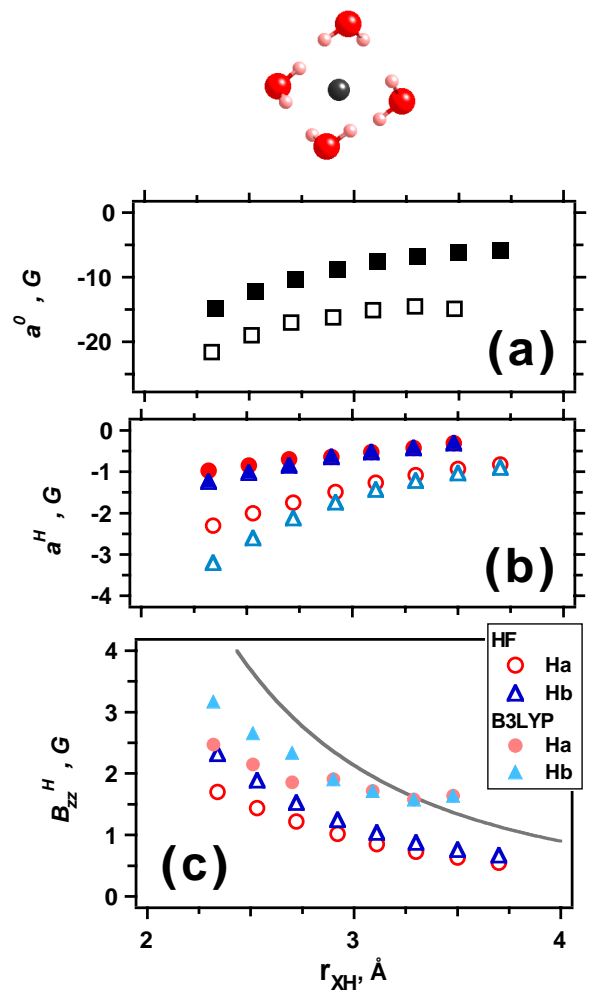

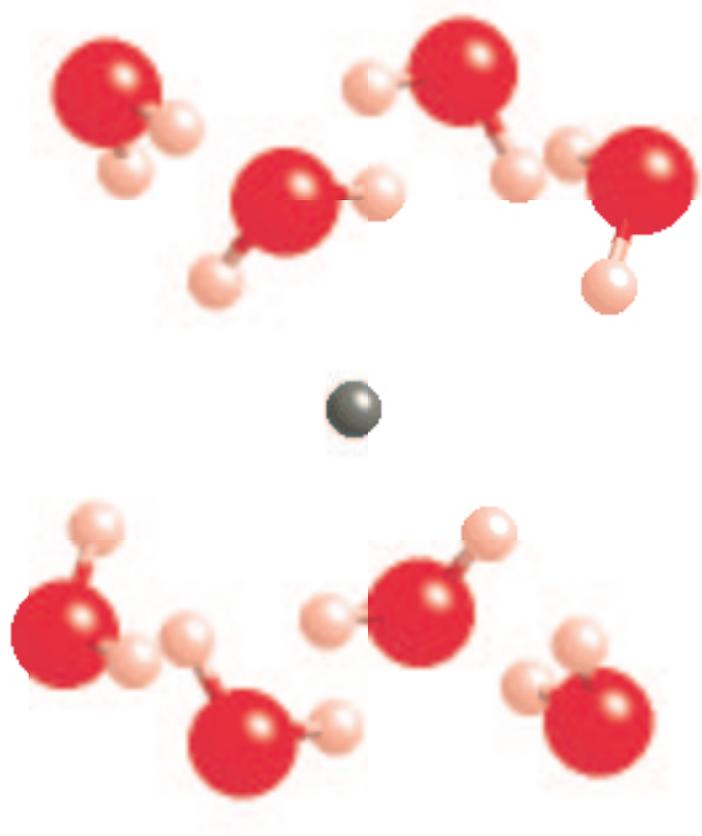

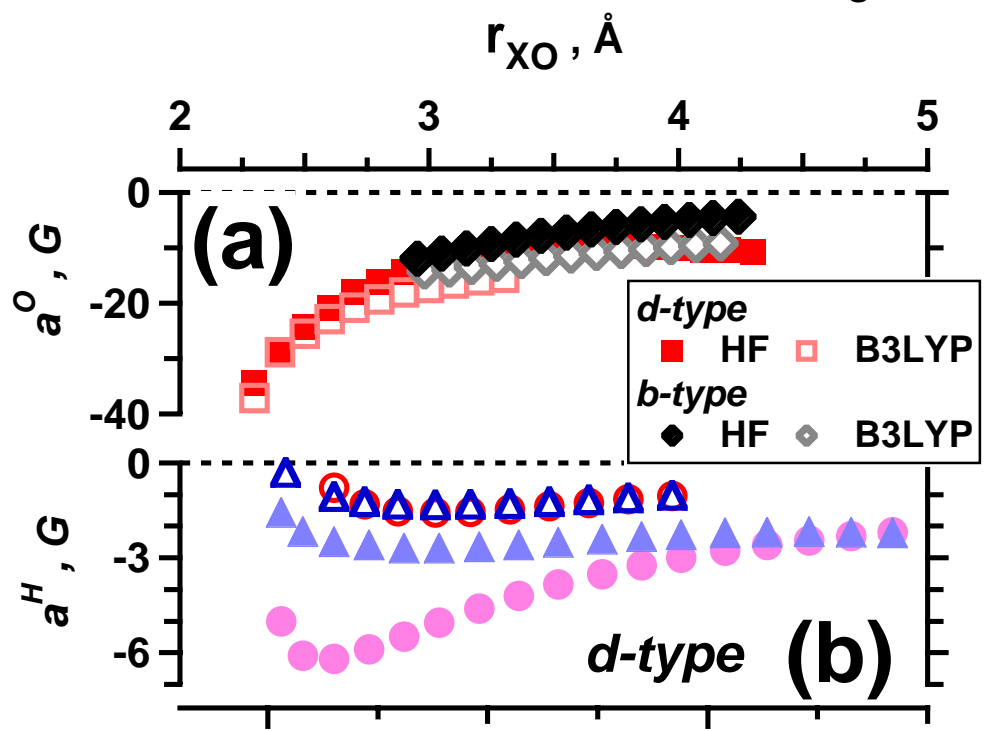

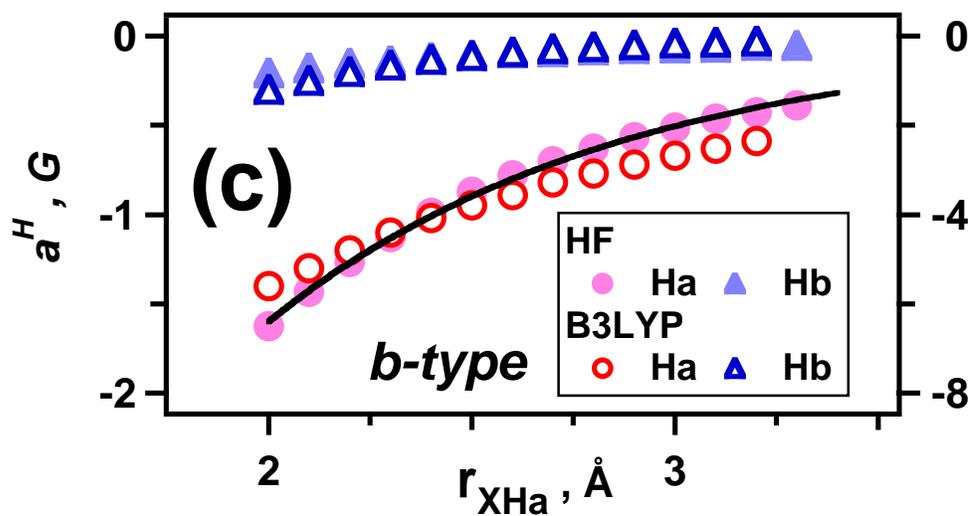

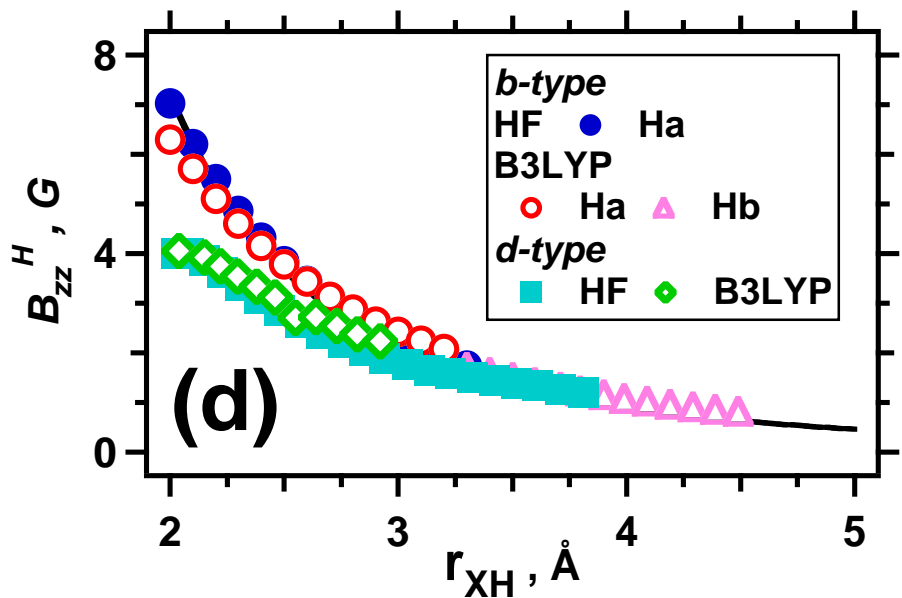

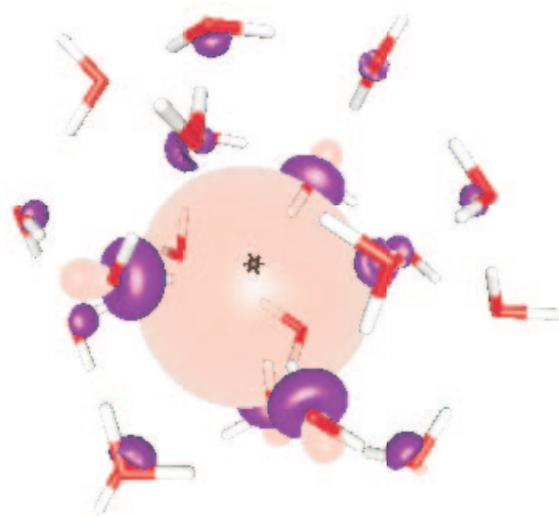

**(a)** ±**0.03**

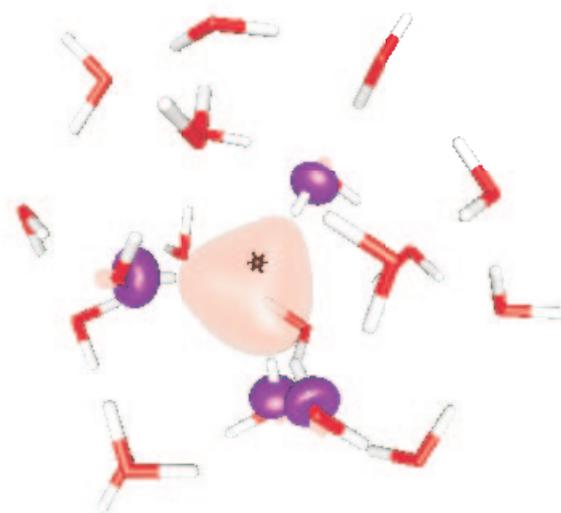

**(b)** ±**0.06**

**w20n2-**

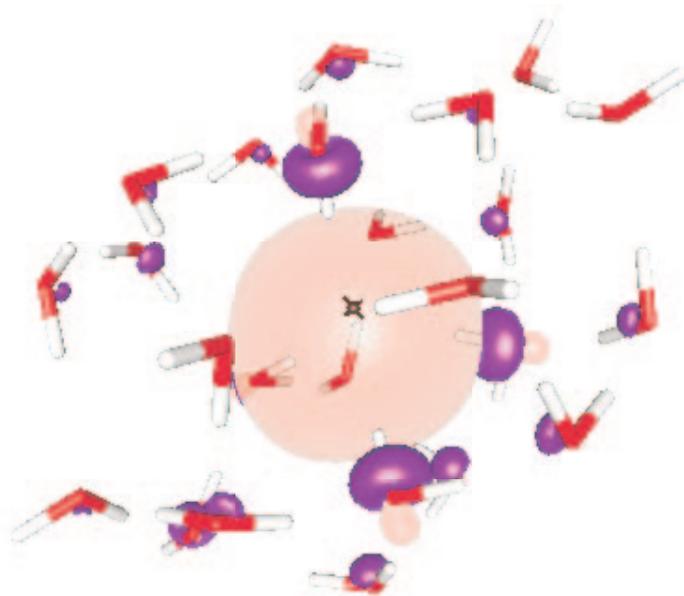

**(a) ±0.03**

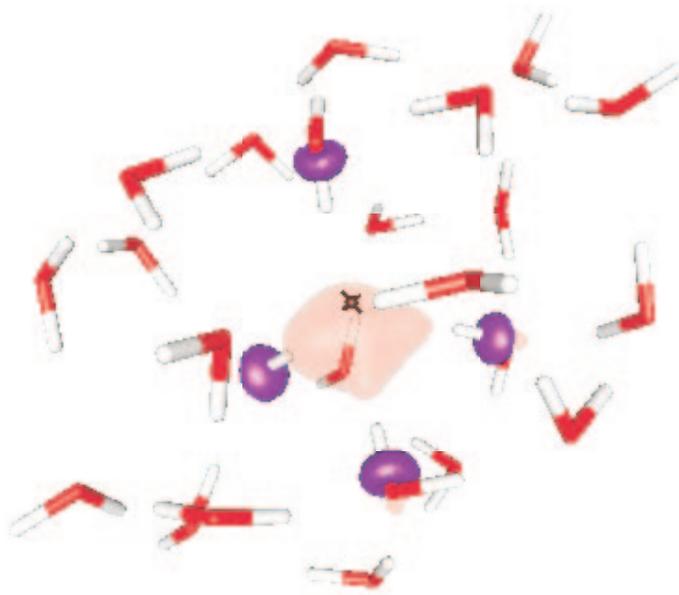

**(b) ±0.06**

**t24n1-**

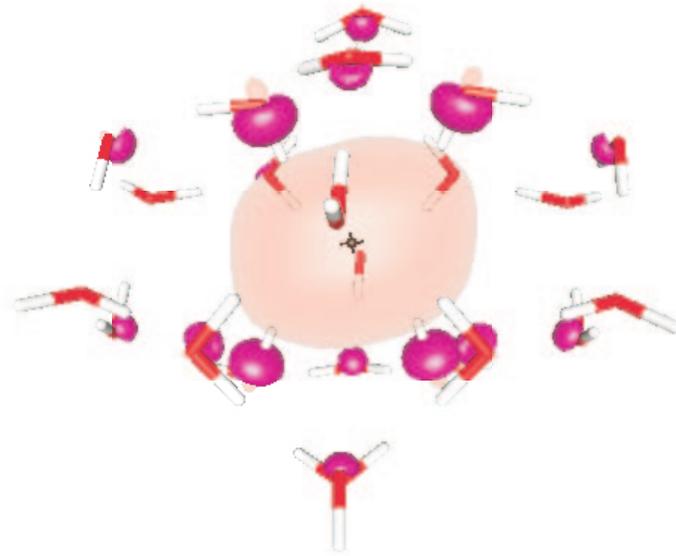

**(a)** ±**0.03**

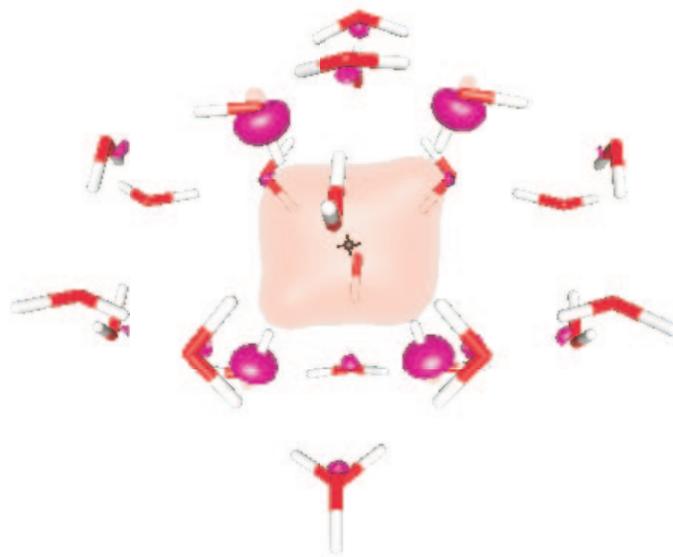

**(b)** ±**0.04**

$4^6 6^8 B^-$

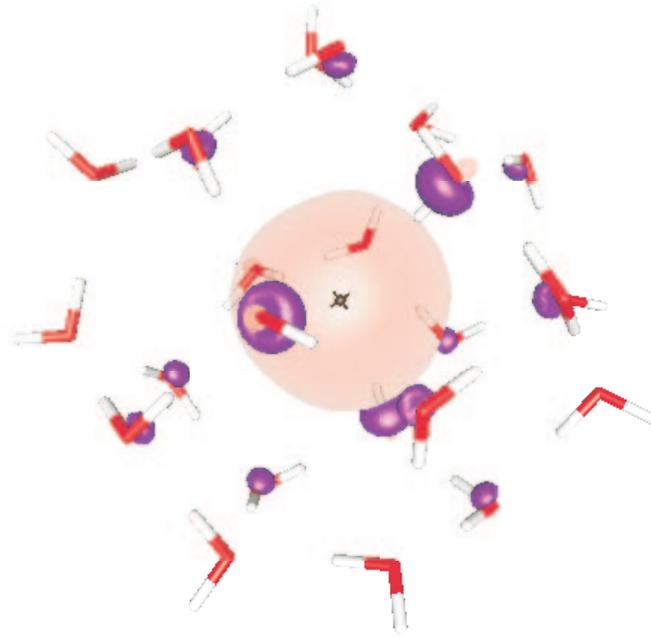

**(a)** ±0.03

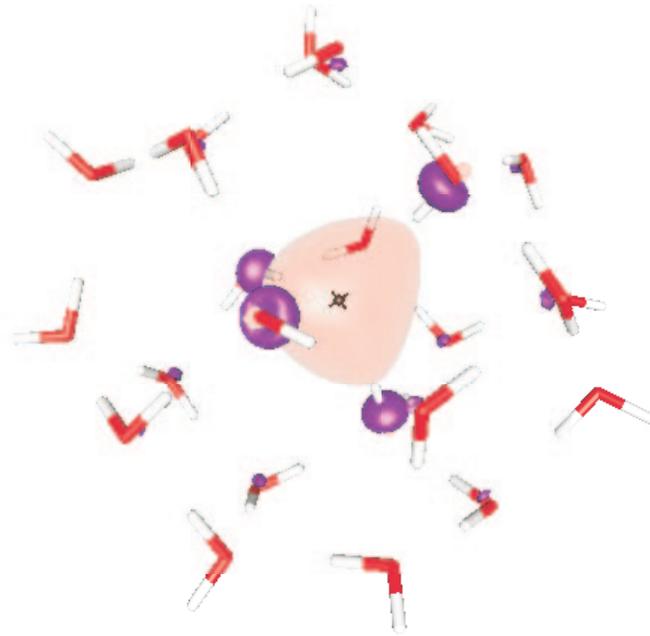

**(b)** ±0.04

$5^{12}6^{2}B-$